\documentclass[preprint]{mn2e}
\usepackage{epsfig}

\newcommand{\be}{\begin{displaymath}}
\newcommand{\ee}{\end{displaymath}}
\newcommand{\bea}{\begin{eqnarray}}
\newcommand{\eea}{\end{eqnarray}}

\title[Stellar mass black hole binaries as ULXs]
{Stellar mass black hole binaries as ULXs}

\author[S. Rappaport, Ph. Podsiadlowski and E. Pfahl]
{S.\ A.\ Rappaport$^{1}$, Ph.\ Podsiadlowski$^{2}$ and E. \ Pfahl$^{3}$\\
\it
$^1$ {Department of Physics, MIT 37-602B, 77 Massachusetts Ave, 
                 Cambridge, MA 02139 (\tt sar@mit.edu})\\
$^2$ {Department of Physics, Oxford University, Oxford, UK
(\tt podsi@astro.ox.ac.uk})\\
$^3$ {Chandra Fellow; Harvard-Smithsonian Center for Astrophysics; 
60 Garden St., Cambridge, MA 02138 (\tt epfahl@cfa.harvard.edu)}}

\date{\today}
\volume{000}
\pagerange{\pageref{firstpage}\,--\,\pageref{lastpage}}\pubyear{2004}
\begin{document}
\maketitle
\label{firstpage}
\begin{abstract}

Ultraluminous X-ray sources (ULXs) with $L_x > 10^{39}$ ergs s$^{-1}$
have been discovered in great numbers in external galaxies with {\em
ROSAT}, {\em Chandra}, and {\em XMM}.  The central question regarding
this important class of sources is whether they represent an extension
in the luminosity function of binary X-ray sources containing neutron
stars and stellar-mass black holes (BHs), or a new class of objects,
e.g., systems containing intermediate-mass black holes (100\,--\,1000
$M_\odot$).  We have carried out a theoretical study to test whether a
large fraction of the ULXs, especially those in galaxies with recent
star formation activity, can be explained with binary systems
containing stellar-mass black holes.  To this end, we have applied a
unique set of binary evolution models for black-hole X-ray binaries,
coupled to a binary population synthesis code, to model the ULXs
observed in external galaxies.  We find that for donor stars with
initial masses $\ga 10~M_\odot$ the mass transfer driven by the
normal nuclear evolution of the donor star is sufficient to
potentially power most ULXs.  This is the case during core hydrogen
burning and, to an even more pronounced degree, while the donor star
ascends the giant branch, though the latter phases lasts only
$\sim$5\% of the main sequence phase.  We show that with only a modest
violation of the Eddington limit, e.g., a factor of $\sim$10, both the
numbers and properties of the majority of the ULXs can be reproduced.  One 
of our conclusions is that if stellar-mass black-hole binaries account for 
a significant fraction of ULXs in star-forming galaxies, then the rate of
formation of such systems is $\sim 3 \times 10^{-7}$ yr$^{-1}$ 
normalized to a core-collapse supernova rate of 0.01 yr$^{-1}$.

\end{abstract}


\section{Introduction}

The {\em Chandra X-ray Observatory} and the {\em XMM} mission have
been used to study entire populations of accretion powered binary
X-ray sources in external galaxies.  At distances exceeding $\sim$10
Mpc, the sources that can be studied are limited to the luminous end
of the distribution function, e.g., $L_x \ga 10^{37}$ ergs
s$^{-1}$.  The observed sources with $L_x$ up to a few
$\times~10^{38}$ ergs s$^{-1}$ are very likely closely related to the
high- and low-mass X-ray binaries that have been well studied for the
past four decades in our own Galaxy and its neighbors.  However, the
discovery of ultraluminous X-ray sources (ULXs) with {\em Einstein}
(Fabbiano 1989), {\em ROSAT} (Colbert \& Ptak 2002; Roberts \&
Warwick 2000), and {\em ASCA} (Makashima et al.\ 2000) has been
greatly extended by {\em Chandra} and {\em XMM} with their far
superior sensitivity (see, e.g., reviews by Fabbiano \& White 2004;
Colbert \& Miller 2004).  These sources are typically defined to have
$L_x \ga 10^{39}$ ergs s$^{-1}$ ($2-10$
keV) and have been observed to luminosities as high as a few
$\times~10^{40}$ ergs s$^{-1}$.  A key question which observations of
these sources seek to answer is whether the compact object is (1) a
neutron star of mass $\sim 1.4~M_\odot$ or black hole of up to $\sim
15~M_\odot$ (see, e.g., Tanaka \& Lewin 1995; Greiner et al.\ 2001;
Lee et al.\ 2002; McClintock \& Remillard 2004), or (2) a black hole of
``intermediate mass'', e.g., $100-1000~M_\odot$ (e.g., Colbert \&
Mushotzky 1999).  This is the question we address on theoretical
grounds in the current work.

ULXs appear in different types of galaxies, including ellipticals
(Angelini, Loewenstein, \& Mushotzky 2001; Jeltema et al.\ 2003) where
their luminosities are generally confined to $L_x \la 2 \times
10^{39}$ ergs s$^{-1}$ (Irwin, Bregman, \& Athey 2004).  ULXs,
a few with luminosities as high as $\sim 5 \times 10^{40}$
ergs s$^{-1}$, are especially prevalent in galaxies with starburst
activity, including ones that have likely undergone a recent dynamical
encounter (e.g., Fabbiano, Zezas, \& Murray 2001; Wolter \& Trinchieri
2003; Belczynski et al.\ 2004; Fabbiano \& White 2004; Colbert \&
Miller 2004).  Two highly photogenic examples are the Antennae and
Cartwheel galaxies.  The Antennae galaxies (Fabbiano et al.\ 2001)
include 49 very luminous X-ray sources, 18 of which are classified as ULXs
(Zezas et al.\ 2002), and are likely to be by-products of the star formation
triggered by the collision of these galaxies (e.g., Hernquist \& Weil
1993).  Approximately half of the ULXs in the Antennae are identified with 
young star-forming regions while the other half have no apparent counterpart
(see, e.g., Fabbiano \& White 2004).  The Cartwheel galaxy reveals a
substantial number of resolved as well as some unresolved point sources 
in a ring coinciding with starburst activity and punctuated by numerous
HII regions (Wolter \& Trinchieri 2003, 2004; Gao et al.\ 2003).  This ring
of star formation is apparently propagating outward at $\sim 50$ km
s$^{-1}$, and the original disturbance was presumably triggered by the
penetration of a smaller galaxy some $5 \times 10^8$ years ago.  The
{\em Chandra} sensitivity limit at the distance of the Cartwheel
($\sim$ 120 Mpc) is $L_x \simeq 5 \times 10^{38}$ ergs s$^{-1}$.

The ULXs found in these galaxies and numerous others have been suggested to harbor ``intermediate-mass black holes'' (IMBHs, e.g., Colbert \& Mushotzky 1999).  The motivation 
for this is clear. The Eddington limit for spherically symmetric accretion is given by 
\bea
L_{\rm Edd} \simeq 2.5 \times 10^{38} \frac{M}{M_\odot} (1+X)^{-1} 
~~~ {\rm ergs~ s}^{-1}  ~~~,
\eea
where $M$ is the mass of the accretor, $X$ is the hydrogen mass
fraction in the accreted material, and Thomson scattering is taken to
be the dominant source of opacity.  The Eddington limit for neutron
stars is only $\sim 2 \times 10^{38}$ ergs s$^{-1}$, though a few
accretion-powered X-ray sources have typical persistent luminosities
of $\sim 5-8 \times 10^{38}$ ergs s$^{-1}$ (e.g., Levine et al.\
1991, 1993)---not quite in the ULX range.  Intermediate-mass black holes, by
contrast, would have Eddington luminosities of $\sim 10^{40} -
10^{41}$ ergs s$^{-1}$ which nicely cover the ULX range.  Moreover,
the expected spectra from intermediate-mass black holes accreting
substantially below their Eddington limit (as would be the case for
most of the ULXs if they were IMBHs), would have low inner-disk
temperatures, as is inferred for some of the ULXs (Miller et al.\
2003; Miller, Fabian, \& Miller 2004a,b; Cropper et al. 2004).  However, 
before invoking a new type of hitherto unobserved object, it is sensible to ascertain
whether stellar-mass black holes of mass $5-15~M_\odot$ could explain
many or most of the ULXs.  The corresponding Eddington limit for these
sources is as high as $\sim 2 \times 10^{39}$ ergs s$^{-1}$, i.e., extends
into the low end of the ULX range.  Several Galactic black-hole
transient X-ray sources have been suspected of exceeding their
respective Eddington limits, but the best case may be GRS 1915+105
where the observed $L_x$ has been above $10^{39}$ ergs s$^{-1}$ about
$\sim$30\% of the time in daily {\em RXTE ASM} averages over the past
8 years (A. Levine, private communication).  On rare occasions, and
for brief intervals, $L_x$ for GRS 1915+105 can be high as $7 \times 10^{39}$ 
ergs s$^{-1}$ ($2-10$ keV; e.g., Greiner et al.\ 1996), which is a few times 
greater than the Eddington limit for the $\sim 14~M_\odot$ black hole in this 
system (Greiner et al. 2001).  If taken at face value, however, sources with 
Eddington limited stellar-mass black holes would still fall somewhat short in 
explaining a significant fraction of the observed ULXs.

A number of ideas have been put forth for ways to circumvent the
problem of how $\sim 10~M_\odot$ black holes could have apparent $L_x$
values as high as a few $\times~10^{40}$ ergs s$^{-1}$.  King et al.\
(2001) suggested that the radiation may be geometrically beamed so
that the true value of $L_x$ does not, in fact, exceed the Eddington
limit.  K\"ording, Falcke, \& Markoff (2002) proposed that the
apparently super-Eddington ULXs are actually emission from microblazer
jets that are relativistically beamed along our line of sight.  However, studies 
of the giant ionization nebulae surrounding a number of the ULXs (Pakull \&
Mirioni 2003) seem to confirm the full luminosity inferred from the
X-ray measurements.  Begelman (2002) and Ruszkowski \& Begelman (2003)
found that in radiation pressure dominated accretion disks
super-Eddington accretion rates of a factor of $\sim$10 can be
achieved due to the existence of a photon-bubble instability in
magnetically constrained plasmas.  They propose that this instability
results in a large fraction of the disk {\em volume} being composed of
tenuous plasma, while the bulk of the {\em mass} is contained in high-density 
regions.  The photons then diffuse out of the disk mostly
through the tenuous regions, thereby effectively increasing the
Eddington limit.  This effect is shown to grow as $M^{1/5}$ and may
yield a super-Eddington factor (hereafter ``Begelman factor'') of
$\sim$10 in disks around stellar-mass black-hole systems.\footnote{We 
note that there are essentially no empirical guidelines available concerning 
the spectral characteristics of a stellar-mass black hole accreting at such 
super-Eddington rates.  They could quite conceivably mimic the spectrum 
of a more massive black hole accreting at sub-Eddington rates.}

In the present work we investigate whether most of the ULX
population---at least in galaxies with current or recent star
formation activity---is consistent with black-hole binaries of
conventional mass with only mildly super-Eddington luminosities.  In
order to carry out this study we combine a unique grid of binary
evolution models with a binary population synthesis code to compute
theoretical X-ray luminosity functions vs. time after a burst of star
forming activity has occurred.  We then test how well these calculated
results match the observations.  To the extent that our models are
successful, we can invert the problem and constrain some of the
uncertain input physics in binary stellar evolution calculations.

In \S 2 we describe how the incipient black-hole binaries are
generated.  We also present a grid of 52 black-hole binary models
evolved through the phase of mass transfer onto the black hole, i.e.,
the X-ray phase.  In \S 3 we discuss how we utilize the population
synthesis tools to generate X-ray luminosity functions for black-hole
binaries as a function of time since a discrete star formation event.  
The results are presented as color images of the evolving luminosity 
function vs. time, as well as quantitative line plots of the populations
vs. X-ray luminosity and vs. time.  In \S 4 we discuss how our results
apply to ULXs, particularly those that are being observed in
increasingly large numbers in external galaxies with {\em Chandra} and
{\em XMM}.

\section{Binary Evolution Calculations}

In our previous study related to ULXs (Podsiadlowski, Rappaport, \&
Han 2003; hereafter PRH) we developed two separate components of the
calculations necessary for a binary population synthesis of these
objects.  In the first, we started with a very large set of massive
primordial binaries and generated a much smaller subset of these that
evolved to contain a black hole and relatively unevolved companion
star.  The product was a set of ``incipient'' black-hole X-ray
binaries with a particular distribution of orbital periods, $P_{\rm
orb}$, donor masses, $M_2$, and black-hole masses, $M_{\rm BH}$, for
each of a number of different sets of input assumptions (see, e.g.,
Fig.\ 2 of PRH).  For this part of the calculation, we employed
various ``prescriptions'', based on single star evolution models for
the primary, simple orbital dynamics associated with wind mass loss
and transfer, assumptions about the magnitude of the wind mass loss
from the primary as well as from the core of the primary after the
common envelope, and natal kicks during the core collapse and
formation of the black hole.  Simple energetic arguments were used to
yield the final-to-initial orbital separation during the common-envelope 
phase wherein the envelope of the primary is ejected.  Here we utilized 
a parameter $\lambda$, which is the inverse of the binding energy
of the primary envelope at the onset of the common-envelope phase in
units of $GM_1M_e/R_1$, where $M_1$, $M_e$ and $R_1$ are the total
mass, envelope mass, and radius of the primary, respectively.  This
parameter strongly affects the final orbital separation after the
common-envelope phase, where smaller values of $\lambda$ correspond to
more tightly bound envelopes, and hence more compact
post-common-envelope orbits.

Conventional energetic arguments for the ejection of the common
envelope yield the following expression for initial--final orbital
separation:
\bea
\left(\frac{a_f}{a_i}\right)_{\rm CE} \simeq \frac{M_c M_2}{M_1}
\left(M_2+\frac{2M_e}{\eta_{\rm CE}\lambda r_{\rm L}}\right)^{-1} ~~,
\eea
(e.g., Webbink 1985; Dewi \& Tauris 2000; Pfahl, Rappaport, \&
Podsiadlowksi 2003), where the subscripts ``1'', ``c'', and ``e''
stand for the progenitor of the black hole, its core, and its
envelope, respectively, and ``2'' is for the progenitor of the ``donor star'' 
in the black-hole system.  The quantity $r_{L}$ is the Roche lobe radius 
of the black-hole progenitor in units of $a_i$, $\eta_{\rm CE}$ is the
fraction of the gravitational binding energy between the secondary and
the core of the black-hole progenitor that is used to eject the common
envelope, and $\lambda$ is defined above.  For typically adopted
parameter values, $\lambda \sim 0.01-1$ (e.g., Dewi \& Tauris 2000),
$\eta_{\rm CE} \simeq 1$, and $r_L \simeq 0.45-0.6$ (for an assumed mass ratio
between the black-hole progenitor and the companion in the range of
$\sim 2:1 \rightarrow 15:1$), the second term within the parentheses
in eq. (2) dominates over the first.  In this case, we find the
following simplified expression for $a_f/a_i$:
\bea
\left(\frac{a_f}{a_i}\right)_{\rm CE} \simeq \frac{r_{L}}{2} \left(\frac{M_c}{M_1M_e}\right) M_2 \lambda \simeq 0.005 \left(\frac{M_2}{M_\odot}\right) \lambda~,
\eea
where the leading factor is $r_{L}/2 \simeq 1/4$, while the factor in
parentheses involving the black-hole progenitor is $\sim 0.020 \pm
0.002~M_\odot^{-1}$ for virtually all of the progenitors we consider.
This explains why the large majority of the incipient black-hole
binaries found by PRH resulted from an initially very wide orbit
($P_{\rm orb} \sim ~ $years---when the primary attains radii of $\sim
1000-2300~R_\odot$) preceding the common-envelope phase in order to
avoid a merger between the secondary and the core of the primary.

The final orbital separations (and corresponding orbital periods)
after the common envelope depend on the initial orbital separations
according to eq. (3).  These, in turn, depend upon a number of issues
associated with the evolution of massive stars and their wind loss
characteristics which were explored in PRH.  We expect, however, that
the distribution of incipient black-hole binaries in the $P_{\rm
orb}-M_2$ plane will lie largely within an envelope bounded at the top
by $P_{\rm orb,t} \propto (a_i M_2 \lambda)^{3/2}$ (motivated by
eq. [3]), and at the bottom by values of $P_{\rm orb,b}$ that just
barely avoid a merger between the core of the black-hole progenitor
and the companion star.  From the results of PRH (see, e.g., their
Figs. 2 \& 3) we can fit the following semi-empirical expressions to
the regions in the $P_{\rm orb}-M_2$ plane where the incipient
black-hole binaries are located:
\bea
P_{\rm orb, t} & \simeq & \frac{5}{2} \left(\frac{M_2}{M_\odot}\right)^{3/2}
 \lambda^{3/2}~~~{\rm d} \\
P_{\rm orb, b} & \simeq & \frac{1}{2} \left(\frac{M_2}{M_\odot}\right)^{1/4}
~~~{\rm d}
\eea
for $M_2 > 1~M_\odot$.  The expression for the lower limit on $P_{\rm
orb}$ results from the radius-mass relation for stars on the main sequence
and the functional dependence of the Roche-lobe radius on mass ratio.  
Furthermore, we find empirically that the lower limit on the donor masses 
in these incipient systems is given by:
\bea
M_{\rm 2,min} \simeq \frac{1}{2\lambda}~M_\odot ~~.
\eea
This expression results from the fact that the product of $M_2
\lambda$ must exceed a certain value (see eq. [3]) in order to avoid a
merger.  A more extensive discussion of the generation of the
incipient black-hole X-ray binaries and the relevant references are
given in PRH.  For other references on this subject see, e.g., King \&
Kolb (1999); Fryer \& Kalogera (2001); Portegies Zwart et al. (1997);
and Nelemans \& van den Heuvel (2001).

\begin{figure} 
\begin{center}
\includegraphics[angle=-90,width=0.47\textwidth]{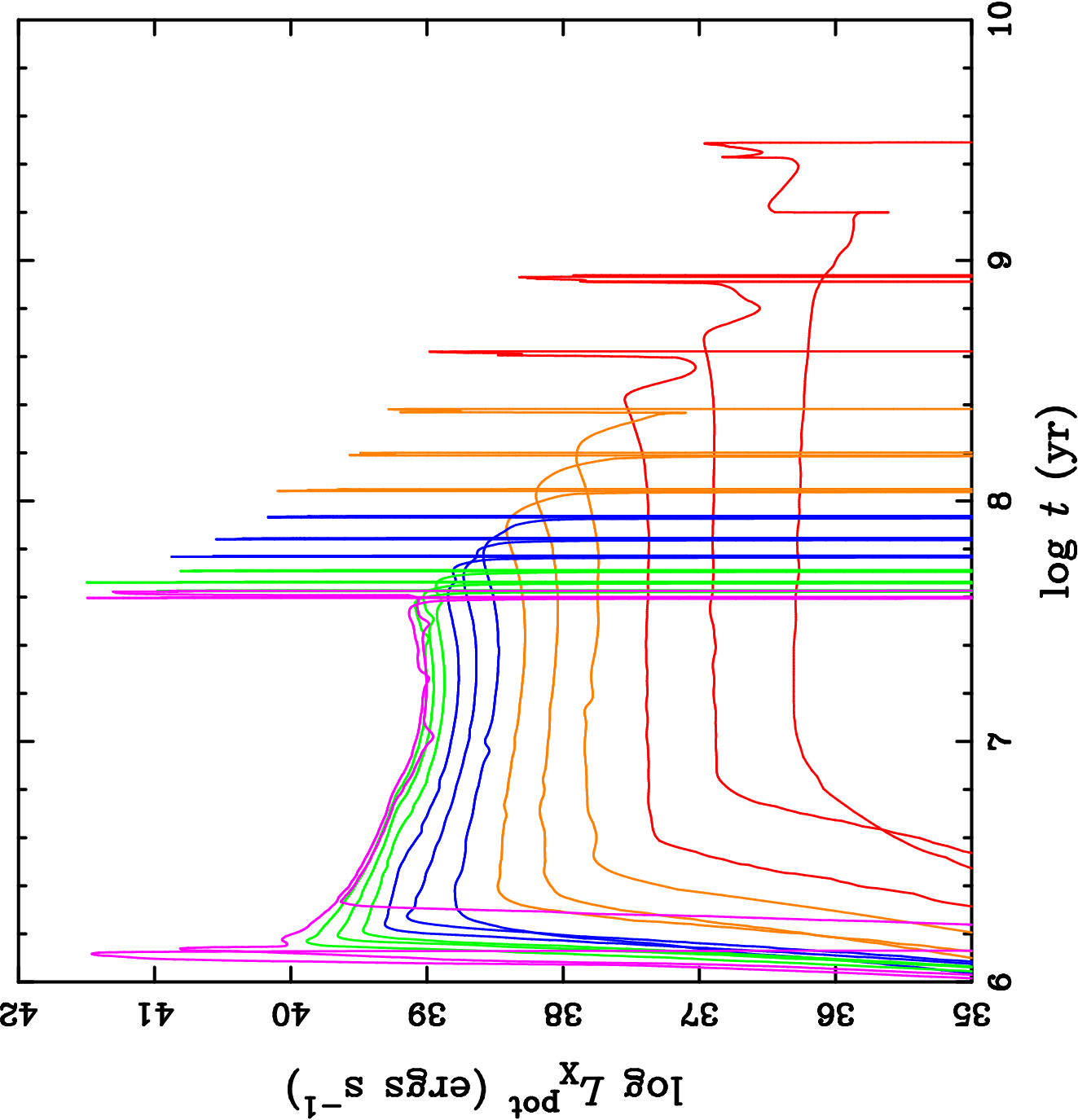}
\caption{Potential X-ray luminosities as a function of time for 14
black-hole X-ray binary evolution sequences; each color corresponds to
a different initial donor star mass ranging from 2 to 17 $M_\odot$
(adapted from PRH).  The trend in the evolutionary sequences from
higher to lower overall X-ray luminosity corresponds to decreasing
initial donor mass. In all cases the donor star is unevolved at the
start of mass transfer, and the black-hole mass is $10 M_\odot$.  The
spiky feature at the end of each evolution corresponds to the donor
star ascending the giant branch.  The duration of this latter phase is
$\sim$5\% of the entire evolution.
\label{fig:14bhev}}
\end{center}
\end{figure}

For the second part of the PRH study we calculated a small but unique
grid of 19 black-hole binary models evolving through the X-ray phase,
i.e., where the donor star transfers mass to the black hole.  These
calculations are done with a full Henyey stellar evolution code so that at 
least the behavior of the donor star throughout its mass-transfer phase 
should be quite accurate and realistic.  Nonetheless, there are still a 
number of uncertainties which involve such issues as whether the 
Eddington limit is strictly adhered to, and when the mass transfer onto the
black hole becomes ``transient'' in nature; we investigate a number of
these issues in this work.  In PRH we systematically explored a subset
of case A binary evolution models where the donor star is initially
unevolved at the time when mass transfer onto the black hole
commences.  Fourteen of the PRH models were for this early case A
binary evolution.  A few additional models were computed for cases
where hydrogen had been partially depleted (also case A) or completely
exhausted (early case B) when mass transfer commenced.  PRH found that
the X-ray lifetime in these latter systems is 1 to 2 orders of magnitudes 
shorter than for systems experiencing case A mass transfer (see Table 1 
of PRH and Figs.\,\ref{fig:4bhev} and \ref{fig:thermal}).  Moreover, 
for large initial mass ratios, mass transfer would be dynamically unstable. 
We therefore expect that these more evolved systems (cases B and C) will 
make only a relatively small contribution to the overall population of ULXs.

\begin{figure}
\begin{center}
\includegraphics[angle=0,width=0.48\textwidth]{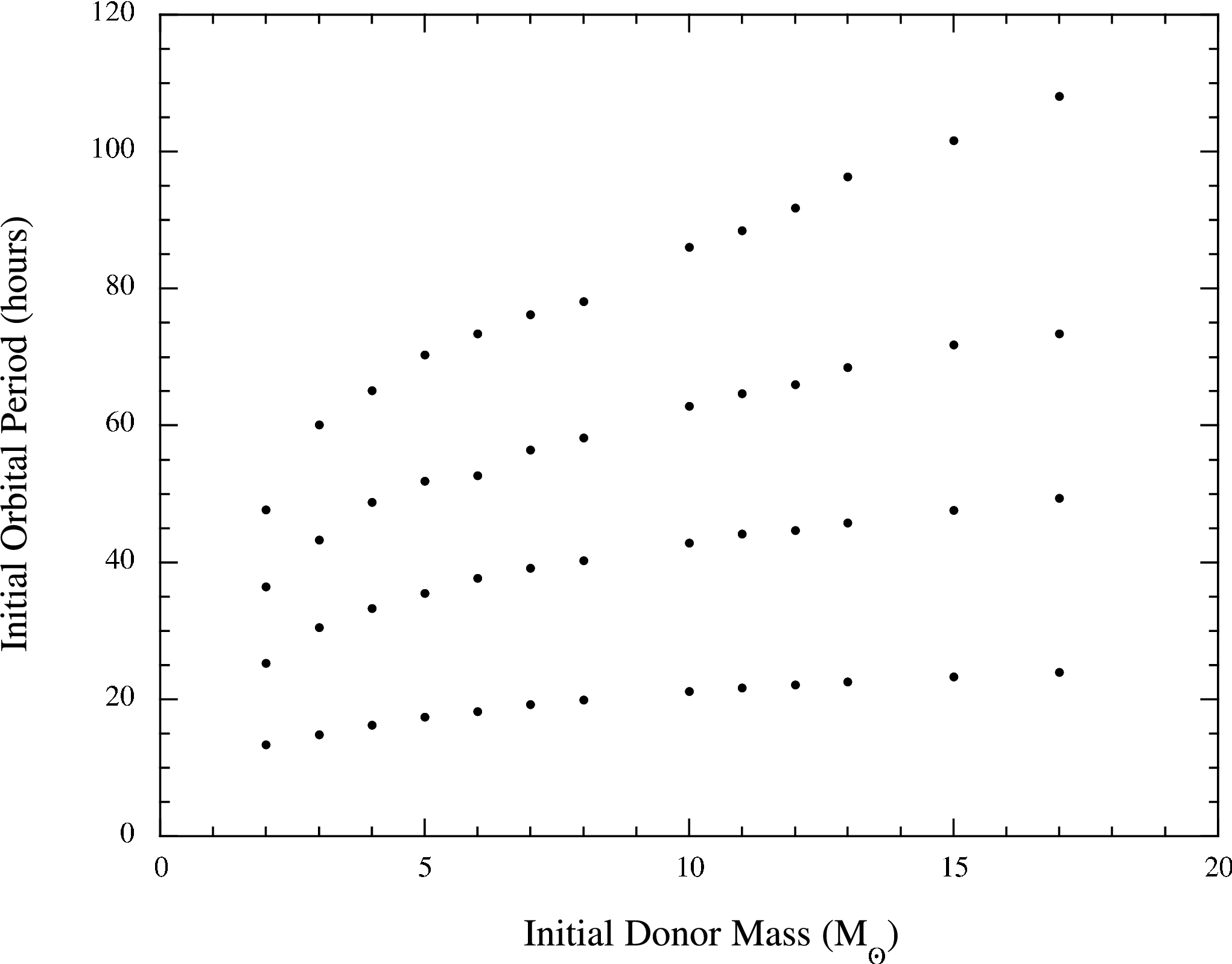}         
\caption{Initial donor masses and orbital periods for the 52
black-hole binary evolution models.
\label{fig:M2vsP}}
\end{center}
\end{figure}

We reproduce in Fig.\ \ref{fig:14bhev} plots of the ``potential''
X-ray luminosity ($L_x^{\rm pot})$ for 14 binary evolution sequences
with initial donor masses of 2 through 17 $M_\odot$ from the original
models presented in PRH (for the case of initially unevolved donors).
The ``potential'' X-ray luminosity refers to the energy output
expected in the absence of the Eddington limit; i.e., if the X-ray
luminosity\footnote{Here, as throughout the paper, all calculated
luminosities refer to the total accretion luminosity without regard
to the region of the X-ray band in which it emerges.  Thus, the $2-10$ 
keV luminosity would likely be somewhat lower than the values cited; 
see \S4 for a discussion of this issue.}, $L_x$, were limited only by the 
mass transfer rate, $\dot M$, and the energy conversion efficiency 
which is dictated by the instantaneous spin of the black hole 
(Bardeen 1970).  We assumed that each black hole starts with a 
mass equal to $10~M_\odot$ and with zero spin (i.e., $j=0$, where 
$0<j<1$ is the dimensionless spin parameter).  We 
reiterate that in each model shown in Fig.\ \ref{fig:14bhev}, the donor 
star is assumed to start on the main sequence when mass transfer 
commences.  We also note that these binary
models include, in addition to $\dot M$ and $L_x$, the evolution of
$P_{\rm orb}$, $M_2$, $M_{\rm BH}$, and the spin of the accreting
black hole.

In the present work we have significantly augmented our grid of black
hole binary evolution models from 19 to 52, and in the process
consider case A mass transfer in a more systematic way.  The
additional models follow the same mass grid as the original 19, except
that for each mass we now allow for three additional evolutionary
states of the donor star at the onset of mass transfer.  This
effectively takes into account a significant portion of the expected
orbital period distribution among the incipient systems.  The
different initial evolutionary states of the donor are characterized
by the hydrogen mass fraction in the core, $X_c$.  Thus, for each
mass, the models are for values of $X_c = 0.7, 0.35, 0.2, {\rm
and}~0.1$.  The corresponding orbital periods at the start of mass
transfer range from about $0.6-4$ days.  The location of the starting
points for these models in the $M_2-P_{\rm orb}$ plane are shown in
Fig.\ \ref{fig:M2vsP}.  To our knowledge, these are the {\em only}
properly computed evolution models for black-hole binaries during
their X-ray phase.

Plots of $L_x^{\rm pot}$ vs. time since the birth of the incipient
binary for a sample of our new evolutionary sequences are shown in
Fig.\ \ref{fig:4bhev}; these are all for the case where the initial
donor mass is $10~M_\odot$, but the initial H fraction is different;
the black, red, blue, and green curves are for $X_c = 0.7, 0.35, 0.2,
~{\rm and}~0.1$, respectively.  Note that, as the donor star is
progressively hydrogen depleted at the start of mass transfer, the
duration of the mass transfer phase becomes systematically shorter,
while the peak values of $L_x^{\rm pot}$ become correspondingly
higher.  The systematically later times of the onset of mass transfer
simply reflect the time that the donor star requires to evolve and
fill its Roche lobe.  The duration of the main mass-transfer phase
becomes shorter, both because the nuclear burning timescale which sets
the rate of evolution of the star becomes shorter (and hence leads to
a higher mass-transfer rate) and because the remaining core
hydrogen-burning lifetime decreases, since less H is available. The
rest of the 52 evolutionary sequences are not shown here in the
interest of space, but are used in the population synthesis study
presented in this work, and are available in digital form upon request
to the authors.
\begin{figure}
\begin{center}
\includegraphics[angle=-90,width=0.47\textwidth]{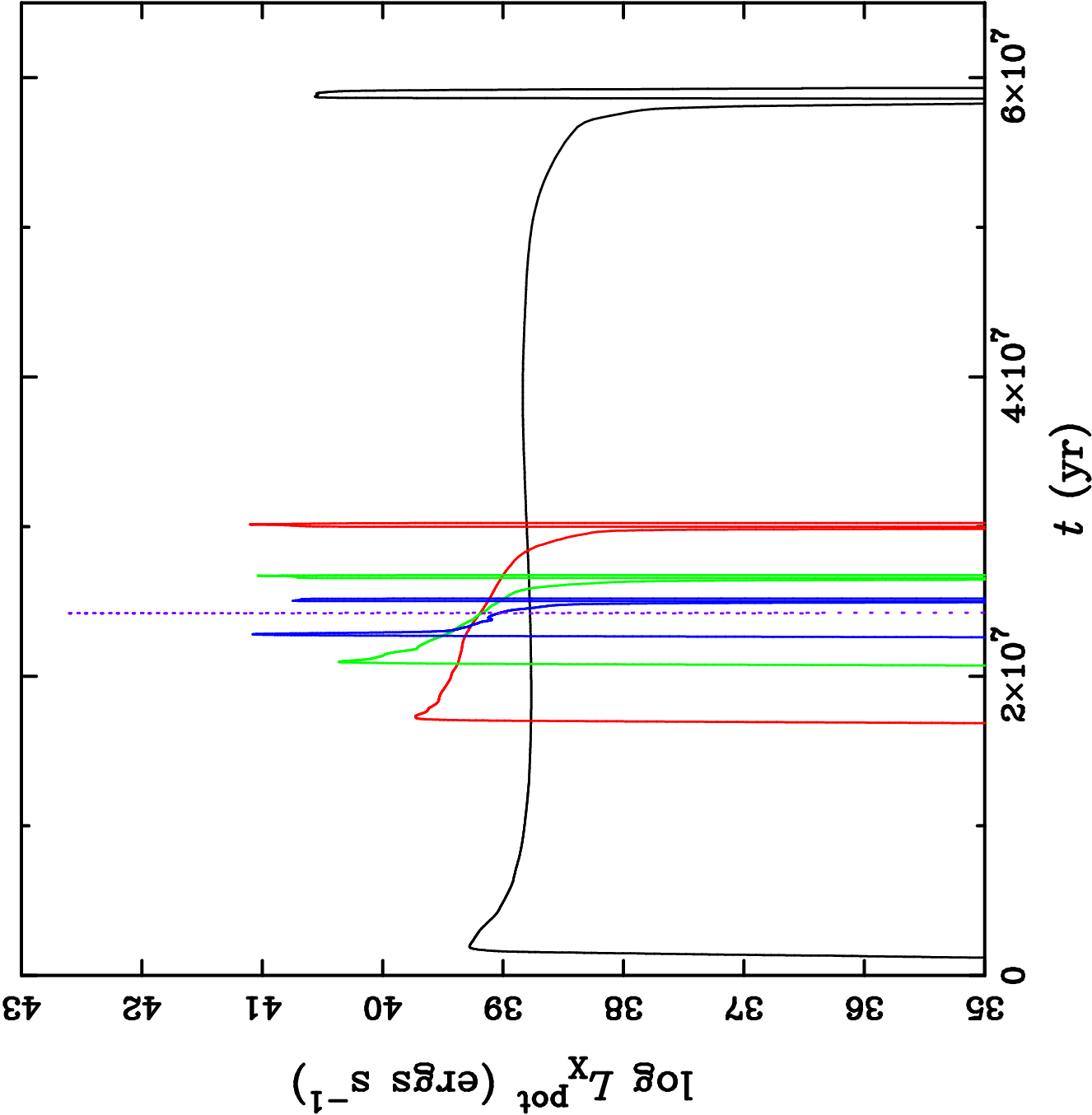}
\caption{Potential X-ray luminosities as a function of time for 5
black-hole X-ray binary evolution sequences; each color corresponds to
a different evolutionary state of the donor star when mass transfer
commences.  In all cases the initial masses of the donor and black
hole are $10\,M_\odot$ and $10\,M_\odot$, respectively.  The black, red,
green, and blue curves correspond to the central hydrogen abundance of
the donor star at the onset of mass transfer of 0.7, 0.35, 0.2, and
0.1, respectively. For comparison, the dotted curve shows the
potential X-ray luminosity for a $10\,M_\odot$ secondary that has already
evolved off the main sequence (so-called early case B mass transfer),
where mass transfer is driven by the expansion of the secondary as it
crosses the Hertzsprung gap on a thermal timescale. The duration of
the mass-transfer phase is not resolved in the figure, as it lasts
only $\sim 50,000\,$yr.
\label{fig:4bhev}}
\end{center}
\end{figure}

Experience has shown that population synthesis calculations for
systems containing collapsed stars typically require $\ga 10^4$
systems in order to achieve results of high statistical quality (see,
e.g., Howell, Nelson, \& Rappaport 2002).  Since each of the 52 models
we have computed required some half hour of cpu time---as well as some
hand holding---it is therefore impractical at this time to run
thousands of black-hole binary models with a Henyey code to describe
the population.  However, we have found from our models that the
evolutions in $L_x$ form a nearly self-similar set.  That is, if we
take any two of the evolutionary models, but especially ones that are
close in initial $M_{2}$ and $P_{\rm orb}$, then the plots of $L_x$
vs. evolution time can be scaled in time and in $L_x$ so they are
nearly the same (see, e.g., Figs. \ref{fig:14bhev} and
\ref{fig:4bhev}).  This is quite different from the case of binary
evolution for systems containing neutron stars (see, e.g.,
Podsiadlowski, Rappaport, \& Pfahl 2002) where the evolution tracks
for initially similar systems can diverge dramatically.  Therefore, we
have made use of this self-similar behavior to develop an
interpolation scheme to produce an effectively much larger set of
models.  Thus, if we wish to know the evolution of a binary system
starting with values of $P_{\rm orb}$ and $M_2$ that are within the
grid boundaries, but are not located at one of the 52 ``nodes'', we
carry out the following variant of a bi-linear interpolation.  First,
a weighting factor for each of the four nearest grid models (in the
$P_{\rm orb}-M_2$ plane) is computed by linearly interpolating in both
$P_{\rm orb}$ and $M_2$.  Once this has been done, the logarithms of
both the beginning and the end times ($t_i$ and $t_f$) of the four
evolutions are interpolated using the weighting factors; this then
provides values of $<t_i>$ and $<t_f>$ to use in the interpolated
evolution.  Each of the four nearest evolutionary models then has its
time axis stretched and shifted so that it begins and ends at $<t_i>$
and $<t_f>$, respectively.  For each time between $<t_i>$ and $<t_f>$
we then do a logarithmic weighting of the quantity of interest (e.g.,
$L_x$) for the four nearest models, using the same weighting factors
as described above.  We have visually inspected a substantial number
of the interpolated evolutions as a check that they behave sensibly.

The black-hole binaries that we produce may account for a significant
fraction of the most luminous X-ray sources born in active star
forming regions.  However, there are several classes of objects not
generated in our population synthesis.  Not included are the compact
black-hole binaries, with low-mass companions, that are transient sources 
and are observed in abundance in our own Galaxy
(see the discussion in PRH).  We argue below that these, in fact, may
be a separate population and not what is being observed in star
formation regions.  Moreover, observations of these sources in our own
Galaxy indicate that they tend to be limited to $\la 10^{38}$
ergs s$^{-1}$, so they may not contribute substantially to the ULX
population.  We do not consider the evolutionary phase before
Roche-lobe overflow when the black hole is fed by accretion from the
stellar wind of the companion.  Such systems are likely to be at the
low end of the luminosity functions we generate in this work, e.g.,
$\sim 10^{34} (M_{\rm BH}/M_\odot)^2$ ergs s$^{-1}$, where $M_{\rm
BH}$ is the mass of the wind-accreting black hole.  Also, we do not
produce systems where the black hole is formed in very wide primordial
binaries without a common envelope being involved, as in the Voss \&
Tauris (2003) scenario.  If objects collapsing to form black holes
have natal kicks, and if these kicks are appropriately scaled down by
the mass of the collapsed star from the case of neutron star kicks
(e.g., to preserve a constant recoil momentum), then the subsequent
orbit could fortuitously remain bound and become highly eccentric.
Eventually, the donor star would evolve to fill its Roche lobe and
mass transfer would then commence.  Such systems will, however, tend
to have much longer orbital periods than the ones we generate.  This
has two consequences: (1) many of these will result in unstable mass
transfer if the donor star has a mass exceeding that of the black hole; 
and (2) the mass-transfer phase will be extremely short lived.  Finally, we
are not considering any systems containing neutron stars since these
are not likely to be able to explain the most luminous ULXs.  On the
other hand, such systems will surely contribute to the luminosity
function up to at least $\sim 5 \times 10^{38}$ ergs s$^{-1}$.
\section{Population Synthesis Calculations}

\begin{figure*}
\begin{center}
\includegraphics[width=0.4\textwidth]{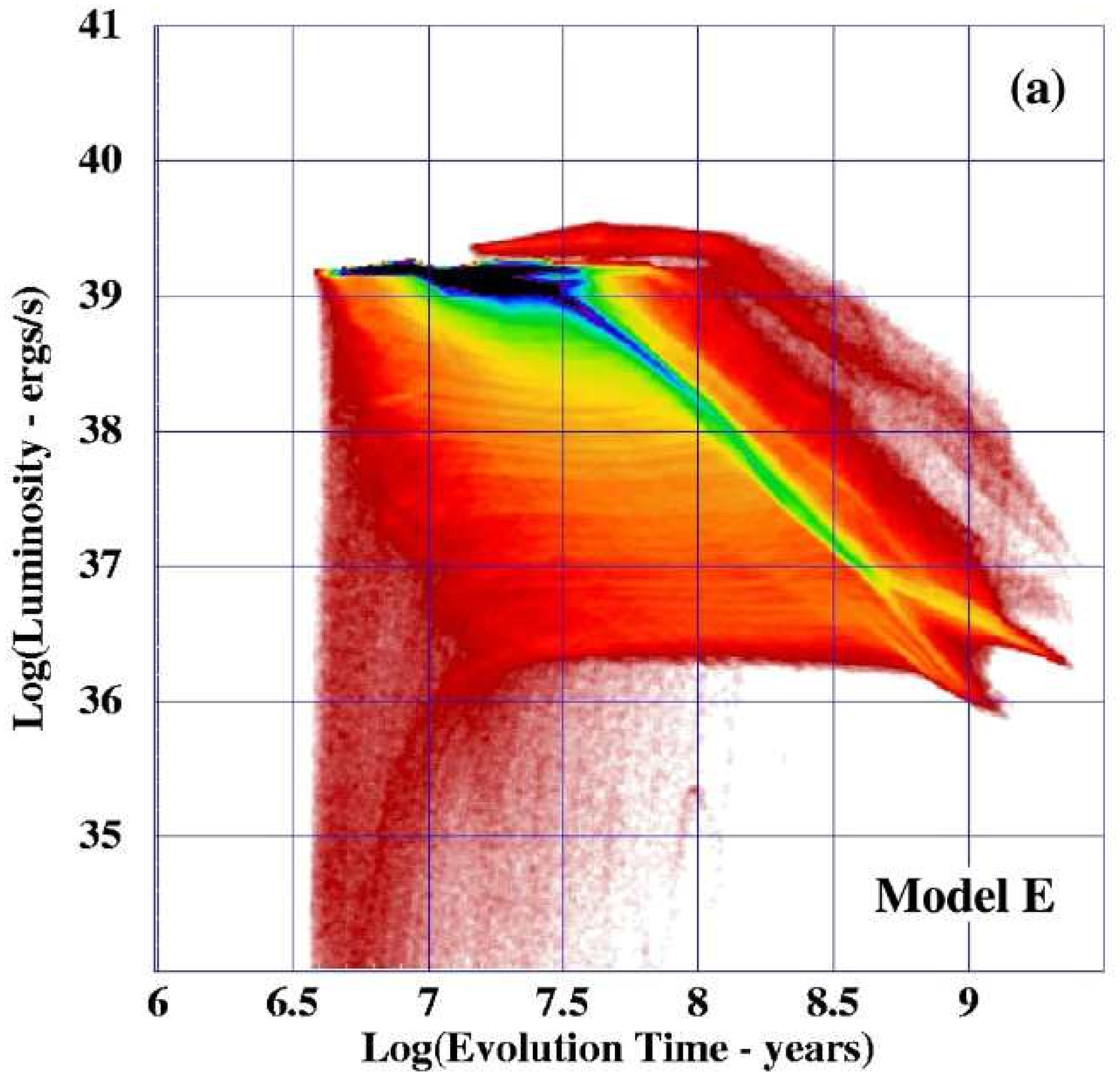}\hglue1cm
 \includegraphics[width=0.4\textwidth]{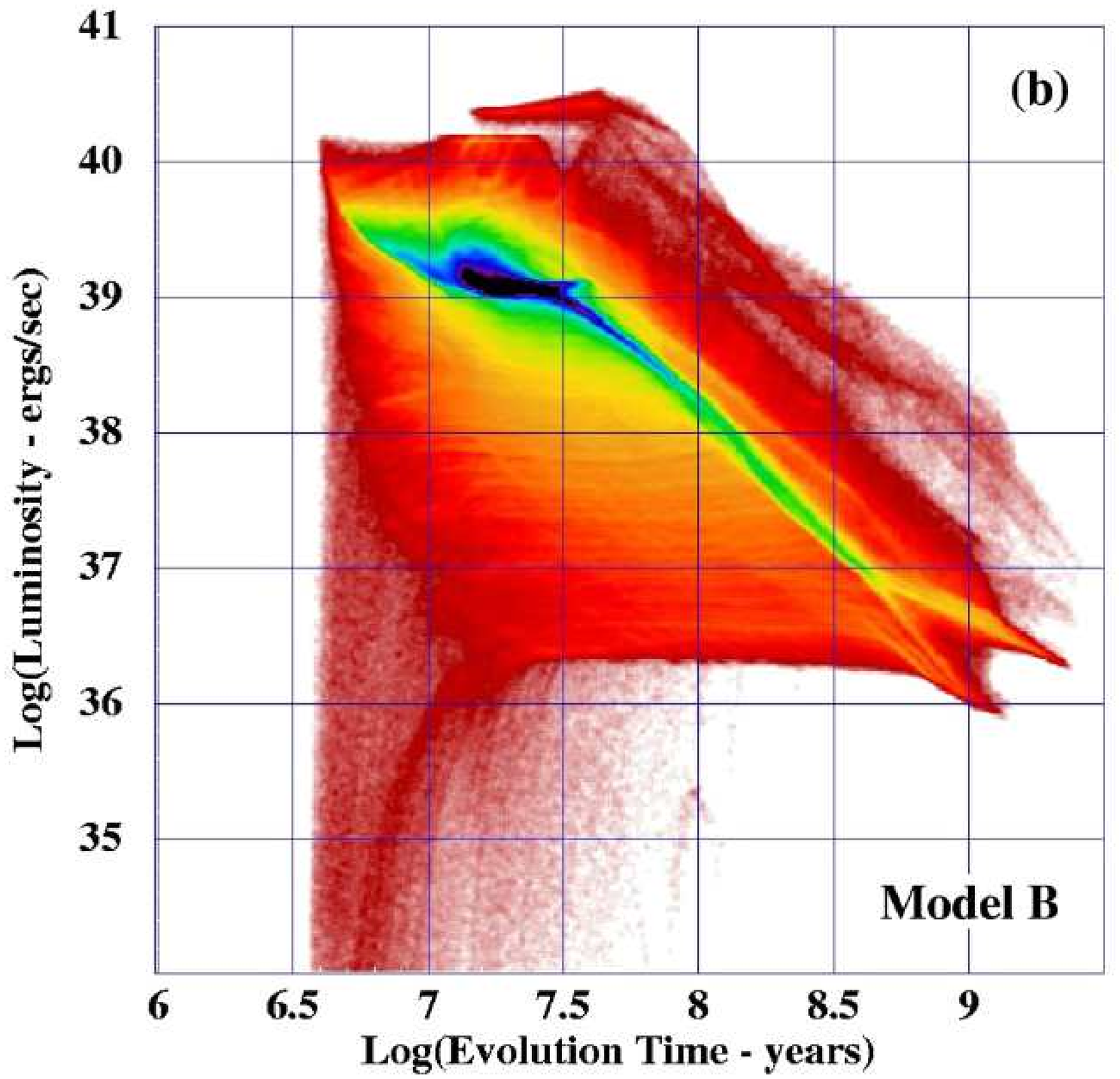}
\vglue1cm
\includegraphics[width=0.4\textwidth]{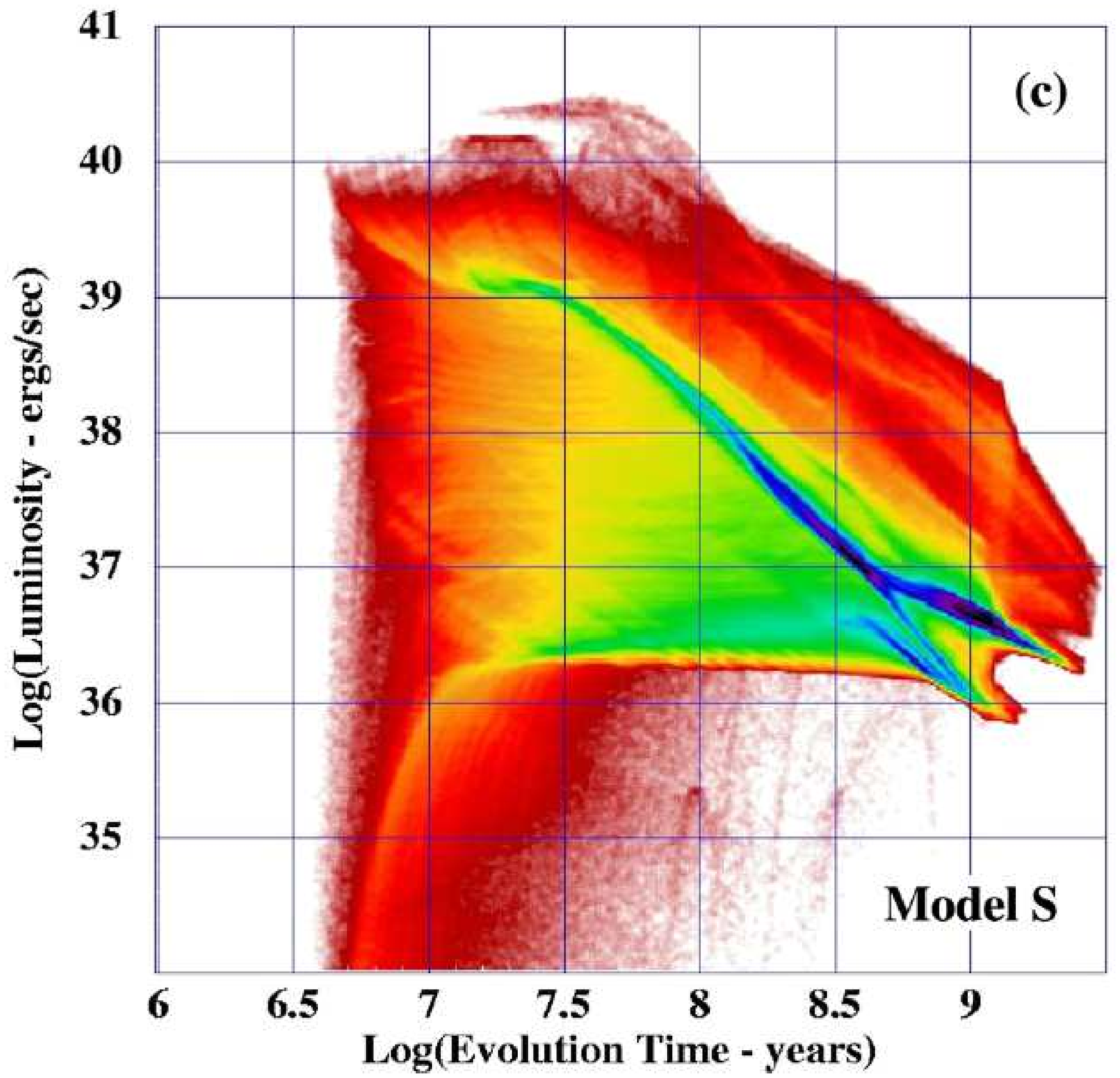}\hglue1cm
\includegraphics[width=0.4\textwidth]{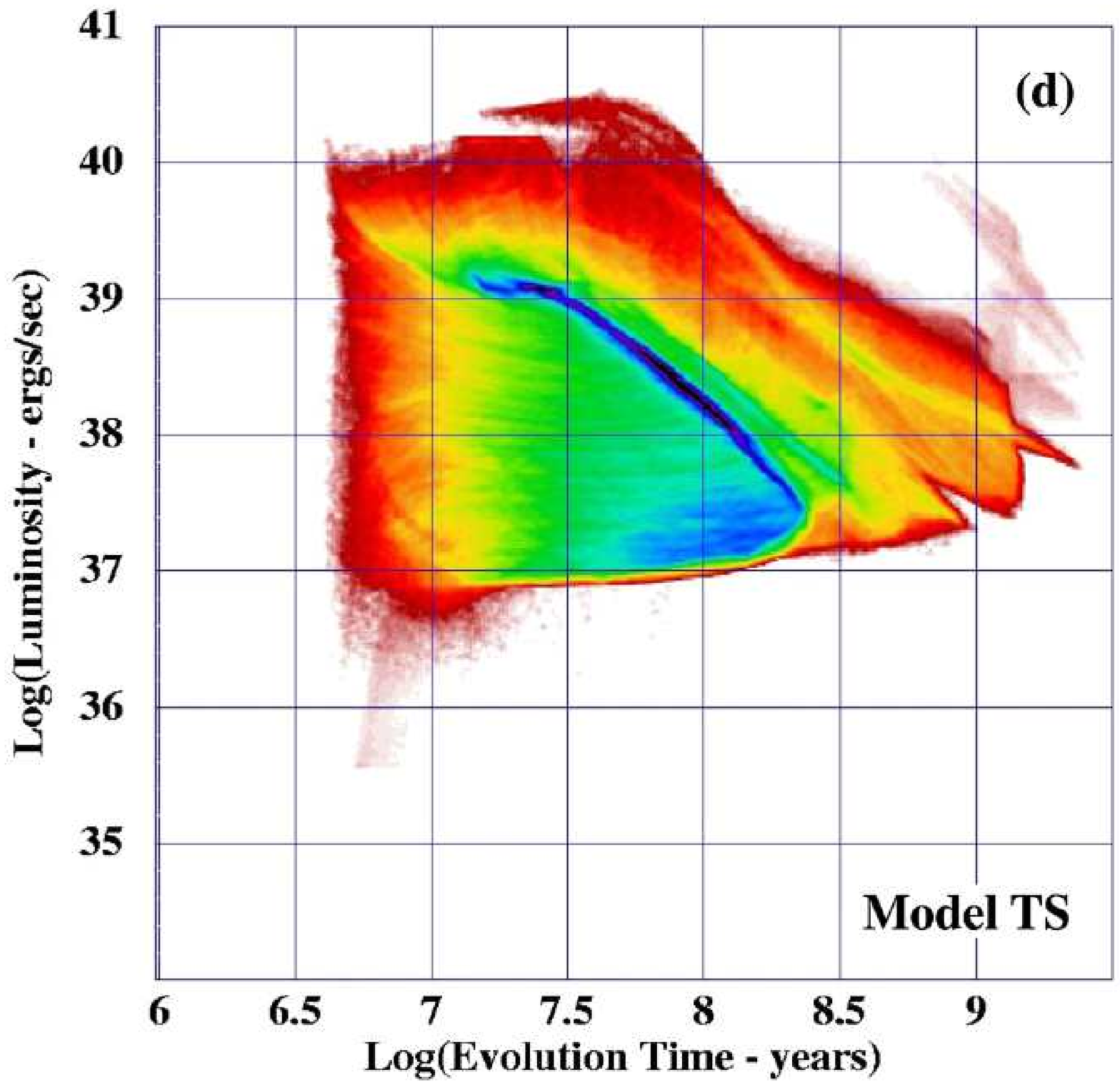}
\caption{Simulated evolution of the X-ray luminosity function of black
hole binaries with time since an impulsive star formation event.
``Luminosity'' refers to the total accretion luminosity without regard
to the region of the X-ray band in which it emerges.  Thus, the $2-10$ keV
luminosity would likely be somewhat lower than the values indicated in the
plots.  Evolution tracks from $10^5$ X-ray sources were computed and then
registered in each of the $700 \times 700$ pixels that are traversed.
The colors crudely represent the logarithm of the relative
populations, with purple through red corresponding to ratios of 200 to
1.  The mass of the black hole at the start of mass transfer is taken
to be $10~M_\odot$.  {\em Panel a: Model E.} The appropriate Eddington
limit of each system has been applied.  The upper red bar-like feature
between $\log t \simeq 7.2-8.2$ represents systems with donor stars on
the giant branch with higher mass transfer rates; the larger
black-hole masses later in the evolution and higher He fraction raise
$L_{\rm Edd}$.  {\em Panel b: Model B.}  All specifications are the same as
for panel (a), except that the Eddington limit is allowed to be
violated by up to a factor of 10 (see text for details), and each 
system radiates the luminosity that corresponds to the mass transfer
rate and the energy conversion efficiency according to the
instantaneous spin of the black hole (Bardeen 1970).  As in panel (a),
the red features at the higher $L_x$ values (at $\log t_{\rm ev} \ga
7.2$) are systems with the donor on the giant branch.  {\em Panel c: Model
S.}  All specifications are the same as for panel (b), except that the
initial mass of the donor star is chosen from a Salpeter initial mass
function (Salpeter 1955).  This approximates the output from our
binary population synthesis code for the case where $\lambda \simeq
0.5$.  Note that compared to panels (a) \& (b) there are many fewer
high $L_x$ systems due to the emphasis on the lower-mass donors.
{\em Panel d: Model TS.}  All specifications are the same as for panel (c),
except that transient behavior of the X-ray sources is approximately
taken into account (see text for details).  Such systems are assumed
to be in an ``on'' state only 3\% of the time, but with a value of
$\dot M$ that is 30 times higher than would be the case if the X-ray
emission were steady.  Note that most of the systems with $L_x
\la 10^{37}$ ergs s$^{-1}$ no longer appear on the figure, but
have been systematically shifted to higher $L_x$, producing the
yellow ridge for main-sequence donors. The faint red feature at
times exceeding $10^9$ yr with high $L_x$, corresponds to transients 
with the donor stars on the giant branch.  \label{fig:4panel}}
\end{center}
\end{figure*}

We now use the tools described above to generate large populations of
black-hole binaries of the type likely to be produced in active star
formation regions.  The first step is to decide which of the outputs
to use from the code that generates the incipient black-hole binaries.
Since there are many uncertainties that go into these calculations, we
have decided to use somewhat more general distributions of incipient
systems in the $P_{\rm orb}-M_2$ plane that are inspired by the
results we found in PRH, rather than taken directly from the specific
output for any particular model.  We utilize three generic types of
distributions for the incipient black-hole binaries.  In the first, we
take $P_{\rm orb}$ to be uniformly distributed between the two curves
given by eqs. (1) and (2), and $M_2$ to be uniformly distributed
between 2 and 17 $M_\odot$.  This allows for the broadest
contributions from all the binary evolution models that we have
calculated, and would directly imply a large value of the parameter
$\lambda$.  The second case we consider is one where $P_{\rm orb}$ is
distributed uniformly as above, but $M_2$ is distributed in the same
way as the stellar initial mass function as deduced by, e.g., Salpeter
(1955), Miller \& Scalo (1979), and Kroupa, Tout, \& Gilmore (1993).
This is a very good approximation to what we find for the high
$\lambda$ case, e.g., $\lambda \ga 0.5$ (PRH).  In this case the
envelope of the primary is relatively easy to eject and secondaries
over a wide range in mass are able to successfully get past this
phase.  Their mass distribution then also roughly resembles that of the
primaries since we utilize a flat mass ratio distribution (see PRH).
The third case we consider is one where $\lambda$ has a small value,
e.g., 0.08.  In this case only the more massive secondary stars are
successful in ejecting the envelope of the primary.  For the specific
value of $\lambda = 0.08$ the incipient population of black-hole
binaries has secondary masses largely confined to $M_2 \ga 6~M_\odot$, 
and the masses are roughly uniformly distributed above this value (see
Fig.\ 2 of PRH).

Throughout this work we have somewhat arbitrarily adopted a fixed black-hole
binary production rate of $\mathcal R_{\rm BH} \simeq 10^{-6}$ yr$^{-1}$
and cite a representative value of $\lambda \simeq 0.1$ in order to 
normalize all of our luminosity functions.  In fact, our earlier binary
population synthesis study (hereafter ``BPS''; PRH) yielded the following
values: $\mathcal R_{\rm BH} \simeq 10^{-6}$ yr$^{-1}$ and $\lambda \simeq 
0.08$ for an incipient black-hole binary population characterized approximately 
by donor masses uniformly distributed between 6 and 17 $M_\odot$, and 
$\mathcal R_{\rm BH} \simeq 3 \times 10^{-6}$ yr$^{-1}$ and $\lambda \simeq 
0.5$ for the case where the donor masses are approximately distributed 
according to the Salpeter  (1955) distribution over the mass range $2-17~
M_\odot$.  On the other hand, no particular BPS model output corresponded
closely to our current choice of a uniform distribution of donor masses between
2 and 17 $M_\odot$, and thus we cannot cite BPS values for $\mathcal R_{\rm BH}$ 
and $\lambda$ for this model.  In all of our BPS models (PRH) the core-collapse 
supernova rate was taken to be $\mathcal R_{\rm SN} = 10^{-2}$ yr$^{-1}$.  We 
comment later in the text on how appropriate we think our adopted rate of 
$10^{-6}$ yr$^{-1}$ is for $\mathcal R_{\rm BH}$.

In addition to the three different distributions of incipient black
hole binaries in the $P_{\rm orb}-M_2$ plane described above, there
are two different assumptions we have made regarding the maximum X-ray
luminosities that can be radiated by a black-hole binary.  In the
first of these, the maximum value of $L_x$ is taken to be just that
given by the Eddington limit which, in turn, is governed by the mass
of the black hole and the hydrogen/helium composition of the accreted
material.  In the second, we allow for the possibility that
luminosities up to 10 times the nominal value of $L_{\rm Edd}$ can be
attained (Begelman 2002; Ruszkowski \& Begelman 2003).  Finally, we
consider one additional case where the luminosity of the X-ray sources
is, under certain physical conditions, transient in its behavior.  The
transient behavior is thought to arise from the well-known
thermal-ionization disk instability (Cannizzo, Ghosh, \& Wheeler 1982;
van Paradijs 1996; King, Kolb, \& Burderi 1996; Dubus, Hameury, \&
Lasota 2001; Lasota 2001).  In particular, if the X radiation from the
central source is not able to maintain a sufficiently high temperature
for the accretion disk (especially for the outer portions), the mass
transport through the disk would be unstable, and vice versa.  In
general, this affects mostly the lower luminosity sources with $L_x
\la 10^{37}$ ergs s$^{-1}$.  The specific prescription we used to
determine whether a particular model at a given time would exhibit
transient behavior is described in detail in PRH (see also \S 4).  When 
one of our sources is in an evolutionary phase where it would be a transient 
we simply allow, in a somewhat {\em ad hoc} manner, the mass transfer
rate onto the black hole to increase by a factor of 30 above the long-term 
average rate of matter flowing into the disk, but then give that source a
probability weighting of only 1/30 in the population, indicating that
it would be in an ``on'' state only $\sim$3\% of the time.  More
empirically correct factors to represent transients might be as low as
0.01 for the duty cycle (McClintock \& Remillard 2004; and references 
therein) and up to 100 times the long-term average mass transfer rate.  
However, as we shall show, transient systems play a relatively minor role 
in our models of the ULX population, and that would not change even if 
we had adopted these latter parameters.
\begin{table*}
\noindent{\bf Table 1. Summary Model Parameters}
\tabcolsep=10pt
\begin{tabular}{cccccc}
\hline\hline
\noalign{\vspace{2pt}}
Name & Model & $N(>10^{39}$\,ergs\,s$^{-1})$ & $N(>10^{39.5}$\,ergs\,s$^{-1})$
& $L_{\rm x,tot}^{(a)}$ & $L_{\rm x,tot}^{(b)}$\\
\noalign{\vspace{2pt}}
\hline
\noalign{\vspace{2pt}}
E & Eddington limited luminosity & 30 & 0.2 & $4 \times 10^{40}$ & $1.5 \times 10^{41}$\\
B & Begelman super Eddington factor & 30 & 6 & $6 \times 10^{40}$ & $2.3 \times 10^{41}$\\
TB & Transient behavior included & 30 & 6 & $6 \times 10^{40}$ & $2.3 \times 10^{41}$\\
S & Salpeter IMF & 4 & 0.7 & $1.0 \times 10^{40}$ & $1.5 \times 10^{41}$\\
TS & Transient behavior included & 4 & 0.7 & $1.0 \times 10^{40}$ & $1.5 \times 10^{41}$ \\
6 & initial donor masses $>6 M_\odot$ & 40 & 8 & $9 \times 10^{40}$ & $2.6 \times 10^{41}$\\
\noalign{\vspace{2pt}}
\hline
\end{tabular}

\medskip (a) $L_{\rm x, tot}$ is in units of ergs s$^{-1}$ and represents the maximum 
in the total X-ray luminosity of a galaxy (due to BH-binaries) after a starburst event 
yielding $10^6$ core-collapse SNe. 
(b) $L_{\rm x, tot}$ is in units of ergs s$^{-1}$ and represents the steady-state
total X-ray luminosity of a galaxy (due to BH-binaries) undergoing $0.1$ core-collapse 
SNe per year.

\end{table*}

In summary, the six models that we have chosen to work with are
defined as follows: \\ 
$\bullet$ Model E is defined to have a uniform incipient population 
in the $P_{\rm orb}-M_2$ plane, and the luminosity during the X-ray 
phase is limited to $L_{\rm Edd}$;\\ 
$\bullet$ Model B is the same, except with a ``Begelman'' 
luminosity enhancement such that $L_x$ is permitted to be as large 
as $10~L_{\rm Edd}$; \\
$\bullet$ Model TB is the same as Model B, but transient source
behavior is also included;\\ 
$\bullet$ Model S is the same as Model B except that $M_2$ is 
distributed according to a Salpeter (1955) mass function rather than 
uniformly in mass; \\ 
$\bullet$ Model TS is the same as Model S, but transient source 
behavior is also included; \\ 
$\bullet$ Model 6 is the same as Model B, except that only values
of $M_2 > 6$ are included.
\begin{figure}
\begin{center}
\includegraphics[angle=-90,width=0.48\textwidth]{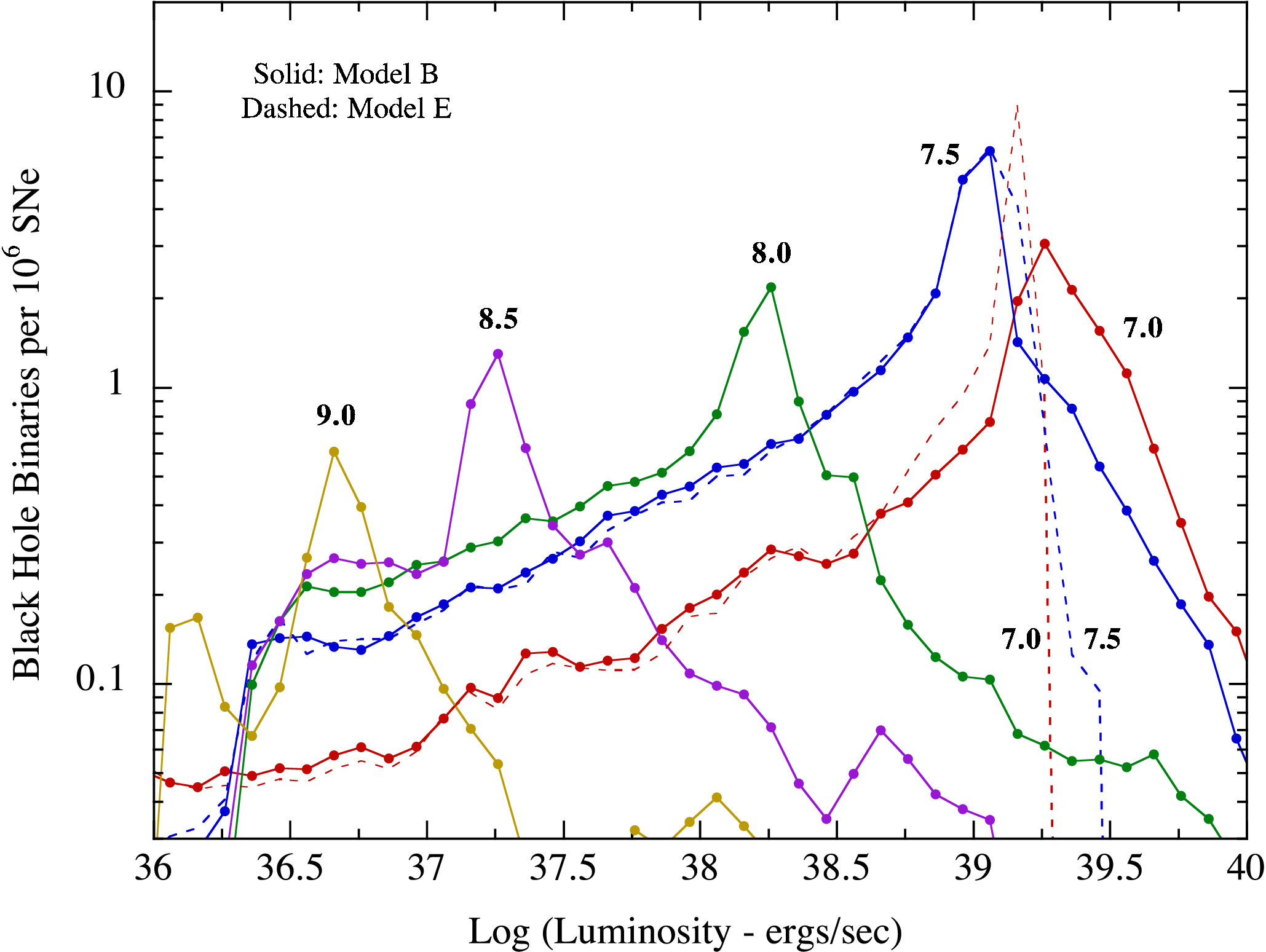}        
\caption{Calculated X-ray luminosity function of black-hole binaries
at five different epochs after an impulsive star formation event
(units are BH-binaries per luminosity bin).  The time labels on each
curve indicate $\log(t_{\rm ev})$ in units of years.  Solid curves are
for Model B (see Fig.\ \ref{fig:4panel}--panel b).  The two dashed
curves are for Model E (i.e., Eddington limited X-ray luminosities;
see Fig.\ \ref{fig:4panel}--panel a) at $\log(t_{\rm ev}) = 7.0$ and
7.5; the curves for all later times are nearly indistinguishable from
those of Model B.  The distributions are normalized to a star
formation event which yielded $10^6$ core-collapse SNe.
\label{fig:lumf}}
\end{center}
\end{figure}

For each of these models we generate a luminosity function vs. time,
$t_{\rm ev}$, since an impulsive star formation event.  We choose, via
Monte Carlo draws, $10^5$ incipient black-hole binaries in the $P_{\rm
orb}-M_2$ plane.  To the extent that a particular value of $P_{\rm
orb}$ and $M_2$ lies within our grid of models, we compute an
interpolated evolution from the 4 nearest models, as described in
detail in \S 2.  To store the results, we set up an array in the $\log
L_x-\log t_{\rm ev}$ plane, containing $700 \times 700$ elements
covering the range $10^{34} < L_x < 10^{41}$ ergs s$^{-1}$ and $10^{6}
< t_{\rm ev} < 10^{9.5}$ yr in equal logarithmic bins.  Each time that
one of our evolution tracks {\rm crosses} an element of this array, a
value of unity is added to that element.  After this operation has
been completed for all $10^5$ interpolated evolution tracks, the
result is displayed as an image representing the {\em unnormalized}
evolving luminosity function for each model.  These are shown for
Models E, B, S, and TS, in Fig.\ \ref{fig:4panel}.  The luminosity
function for Models 6 and TB are not shown in the interest of space.
\begin{figure}
\begin{center}
\includegraphics[angle=-90,width=0.48\textwidth]{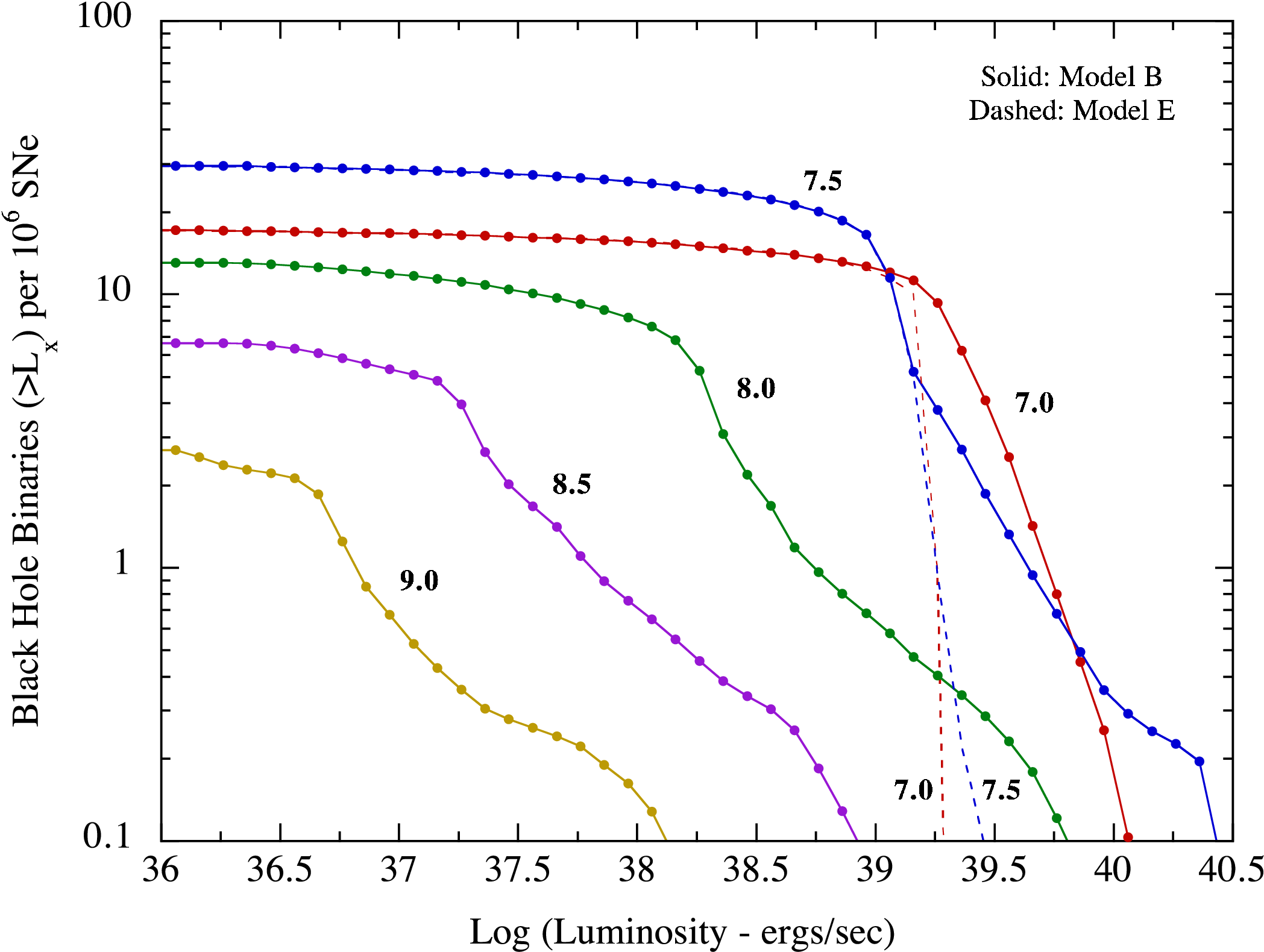} 
\caption{Calculated cumulative X-ray luminosity function of black-hole
binaries at five different epochs after an impulsive star formation
event.  The time label on each curve indicates $\log(t_{\rm ev})$ in
units of years.  Solid curves are for Model B (see Fig.\
\ref{fig:4panel}--panel a).  The two dashed curves are for Model E
(i.e., Eddington limited X-ray luminosities; see Fig.\
\ref{fig:4panel}--panel a) at $\log(t_{\rm ev}) = 7.0$ and 7.5; the
curves for all later times are nearly indistinguishable from those of
Model B.  The normalization of the curves is the same as in Fig.\
\ref{fig:lumf}.
\label{fig:cumlumf}}
\end{center}
\end{figure}

\begin{figure}
\begin{center}
\includegraphics[angle=-90,width=0.48\textwidth]{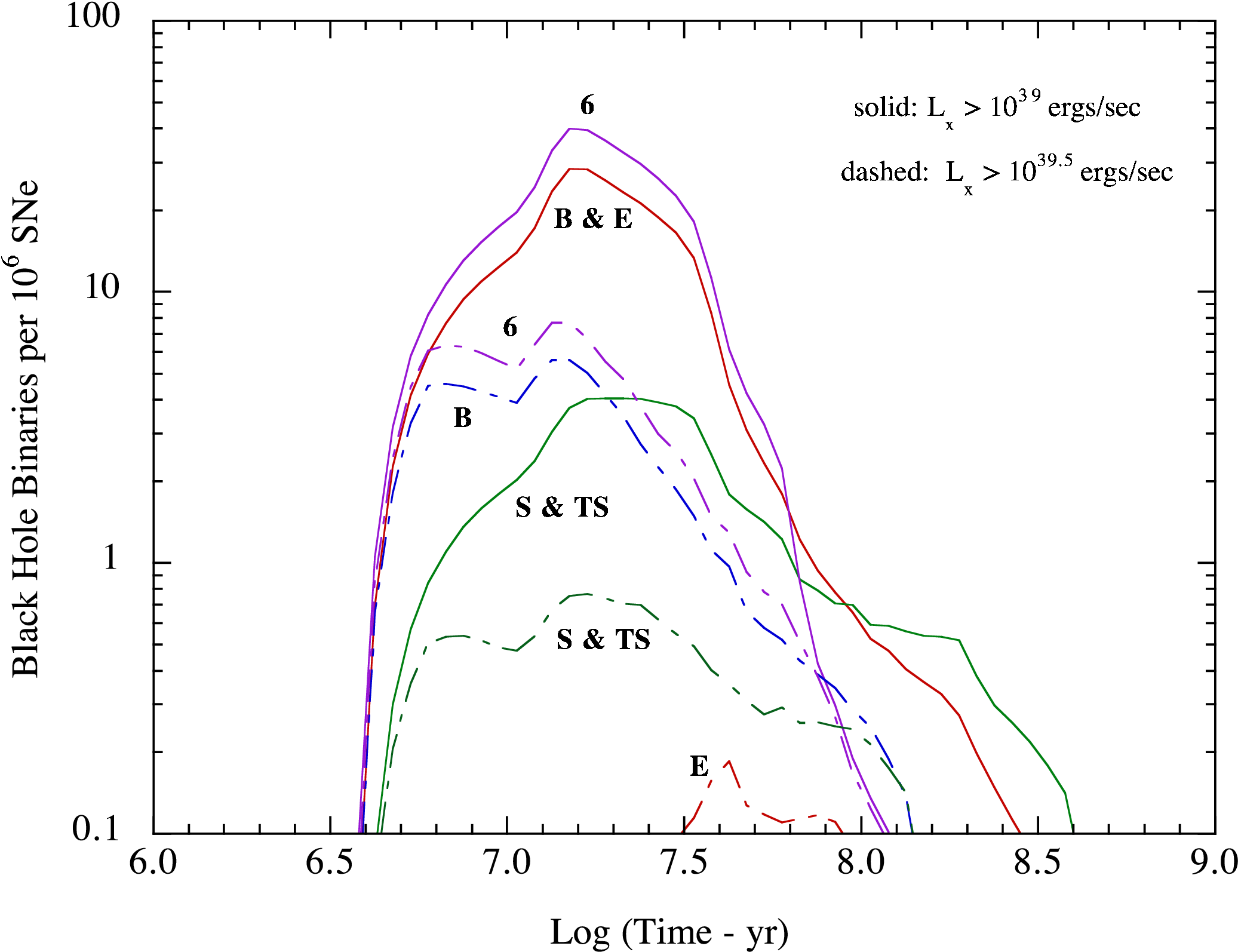}        
\caption{Number of luminous black-hole X-ray binaries as a function of
time after an impulsive star formation event.  Solid curves represent
the number with $L_x >10^{39}$ ergs s$^{-1}$; dashed curves are for
$L_x > 3 \times 10^{39}$ ergs s$^{-1}$.  The normalization of the
curves is the same as in Fig.\ \ref{fig:lumf}.  The peak value for each 
curve is listed in Table 1. \label{fig:nvst}}
\end{center}
\end{figure}

Model E (Fig.\ \ref{fig:4panel}, panel (a)) shows a large
concentration of highly luminous sources ($\ga 10^{39}$ ergs s$^{-1}$;
i.e., ULXs) at $10-30$ Myr.  These result from systems with the
initially more massive companion stars.  Between $\sim$30 and 300
Myr, the peak in the luminosity function falls monotonically from 
$\sim 10^{39}$ to $<10^{37}$ ergs s$^{-1}$ as systems with the more 
massive donors run their evolutionary course, leaving longer-lived, 
lower-luminosity systems with initially lower-mass donors.  Numerous 
systems with $L_x$ exceeding $L_{\rm Edd}$
for a {\em neutron star} are still present after 100 Myr.  Note the
red group of systems at luminosities above the continuous distribution
starting at $\sim$ 15 Myr and continuing to $\ga 1$ Gyr; these
result from the relatively short-lived phase when the donor stars (of
all initial masses) ascend the giant branch and produce mass transfer
rates that increase by 1\,--\,2 orders of magnitude over the preceding
phase of the binary evolution.  Finally, note that the apparent upper
cutoff in the luminosity function seems to increase by about a factor
of 2 for evolution times later than $\sim$ 20 Myr---this results from the 
general increase in the black-hole mass as they grow by accretion, and 
the changing chemical composition of the accreted material which is
becoming He enriched.  

The evolving luminosity function for Model B
(Fig.\ \ref{fig:4panel}, panel (b)) is very similar to that of Model
E, except for the pronounced increase in the allowed upper limit to
$L_x$, by a factor of 10 -- as defined by the model.  This model
allows for a significant population of sources with $L_x$ up to $\sim
3 \times 10^{39}$ ergs s$^{-1}$ and a tailing off population to values
of $L_x \simeq 2 \times 10^{40}$ ergs s$^{-1}$ for times up to 30 Myr
after a star formation event.  We note that the luminosity function
for Model B (Fig.\ \ref{fig:4panel}, panel (b)) extends well past the peak
in the distribution at all times.  Thus, increasing the ``Begelman factor''
by more than a factor of 10 would not substantially enhance the 
population of ULXs.  

In Fig.\ \ref{fig:4panel}, panel (c) we present the evolving
luminosity function for Model S, where the initial donor star masses
are distributed according to a Salpeter IMF, i.e., with a steeply
decreasing population with increasing mass.  In such a scenario, the
X-ray luminosity function is much less strongly peaked at times
$\la 30$ Myr, i.e., not dominated by a narrow range of luminosities, 
than in the previous models.  In particular, there are relatively fewer 
candidate ULX sources.  This is a natural consequence of
the smaller numbers of massive donor stars.  Finally, the evolving
luminosity function for Model TS is presented in Fig.\
\ref{fig:4panel}, panel (d).  In this model, the Salpeter mass
distribution is again applied to the incipient donor stars, but
transient source behavior due to the thermal/viscous disk instability
is now included (in an approximate way).  Note that most of the lower
luminosity sources (i.e., with $L_x \la 10^{37}$ ergs s$^{-1}$)
are missing from the evolving luminosity function.  These have been
boosted up by a factor of 30 in $L_x$, but are given a weighting in
the ``image'' of only 1/30.  The effect of the transients is most
pronounced for times $\ga$200 Myr where they appear as a yellow
ridge in the $L_x-t_{\rm ev}$ plane with $L_x$ in the range of
$10^{38}-10^{39}$ ergs s$^{-1}$.  Note also the faint red feature
appearing at times $\ga 10^9$ yr which extends into the ULX
luminosity range.  These are due to donor stars of initially lower
mass that are on the giant branch and exhibit transient behavior
(GRS 1915+105 in our Galaxy may be an example of such a system).
\begin{figure}
\begin{center}
\includegraphics[angle=-90,width=0.48\textwidth]{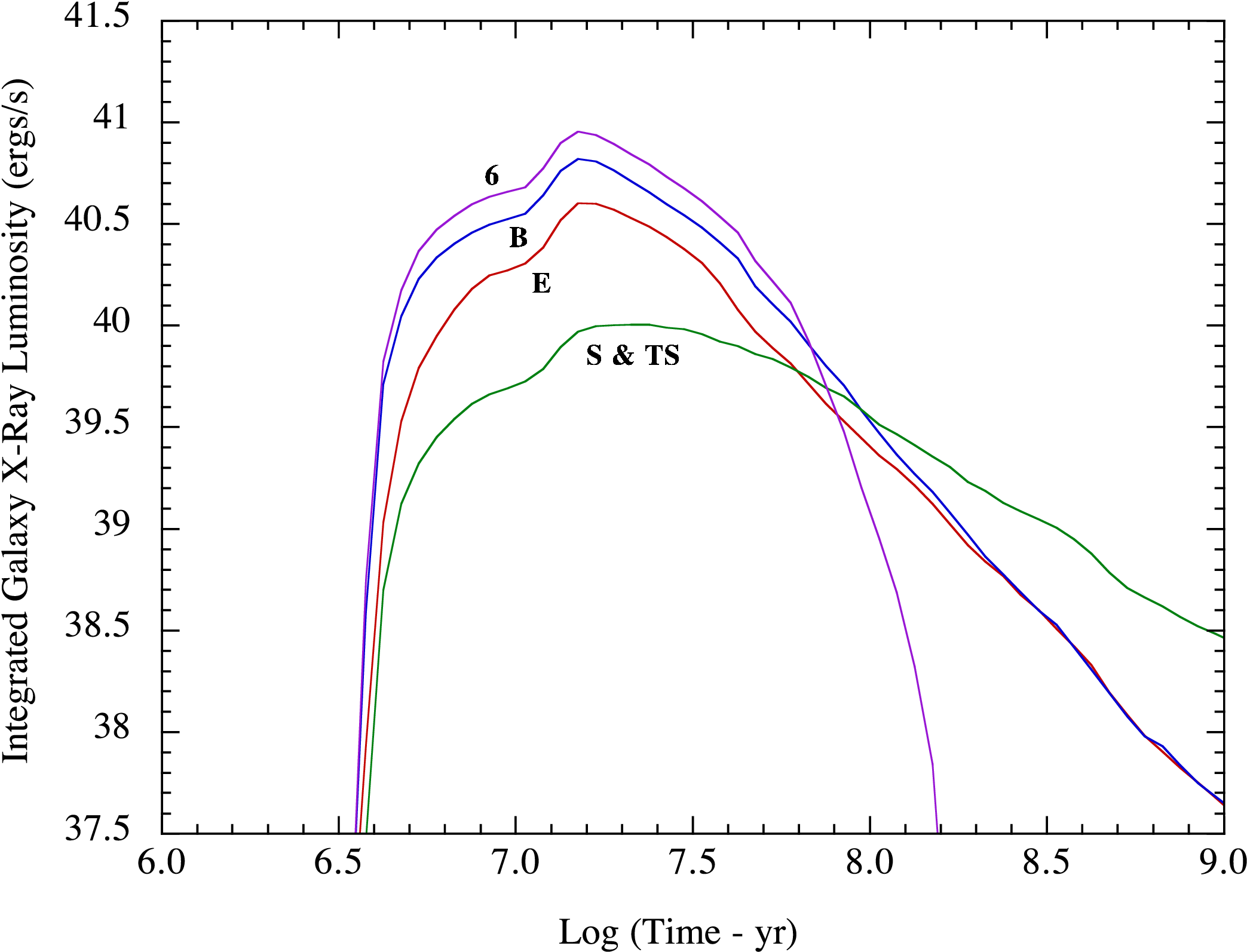}   
\caption{Total luminosity from black-hole X-ray binaries as a function
of time after an impulsive star formation event.  Results for Models
E, B, S, TS, and 6 are shown.  The distributions are normalized to a
star formation event that yielded $10^6$ core-collapse SNe.  The peak
value for each curve is listed in Table 1.
\label{fig:integ}}
\end{center}
\end{figure}

The evolving luminosity functions presented as images are visually
informative and interesting, but a somewhat more quantitative way to
view these distributions is as one-dimensional plots of either the
luminosity functions at specific times or as the evolution with time
of the number of sources exceeding a given $L_x$.  The first of these
is shown in Fig.\ \ref{fig:lumf}.  Here, the luminosity function is
plotted for Model B at 6 different epochs after a star formation
event; these are spaced logarithmically every factor of $\sqrt{10}$ in
time, starting at $10^{7}$ yr.  Also shown for comparison as a set of
dashed lines are luminosity functions for Model E at $\log(t_{\rm ev})
= 7.0$ and 7.5; the curves for all later times are nearly
indistinguishable from those of Model B.  The curves are normalized to
$10^6$ core-collapse supernovae associated with the star formation
event (including those in binary as well as single stars), and to a ratio
of the black-hole production rate to that for core-collapse SNe, 
$\mathcal R_{\rm BH}/\mathcal R_{\rm SN}$, equal to $10^{-4}$.  The 
most prominent appearance of the ULXs, as could be seen qualitatively 
in the luminosity function images, occurs between 10 and 30 Myr.  The
same luminosity functions, but plotted as cumulative distributions,
are shown in Fig.\ \ref{fig:cumlumf}.  From this plot it is easy to
see that between 10 and 30 Myr a galaxy, undergoing sufficient star
formation to produce $10^6$ SNe impulsively (e.g., within $\la
10$ Myr), could be expected to harbor more than a dozen ULXs
containing stellar-mass black holes.  Also, we show in Fig.\
\ref{fig:nvst} the number of ULXs ($\ga 10^{39}$ and $\ga
10^{39.5}$ ergs s$^{-1}$) vs. time since a star formation event for
Models B, E, S, TS, \& 6. These curves have also been normalized to
$10^6$ SNe and to $\mathcal R_{\rm BH}/\mathcal R_{\rm SN} = 10^{-4}$.  
Finally in this regard, we show in Fig.\ \ref{fig:integ} the integrated X-ray 
luminosity corresponding to the source numbers given in Fig.\ \ref{fig:nvst}.  
Except for Model S, typical integrated X-ray luminosities of $4-9 \times 10^{40}$
ergs s$^{-1}$ are attained following a large star formation episode
(normalized to $10^6$ core-collapse SNe).
\begin{figure}
\begin{center}
\includegraphics[angle=-90.,width=0.48\textwidth]{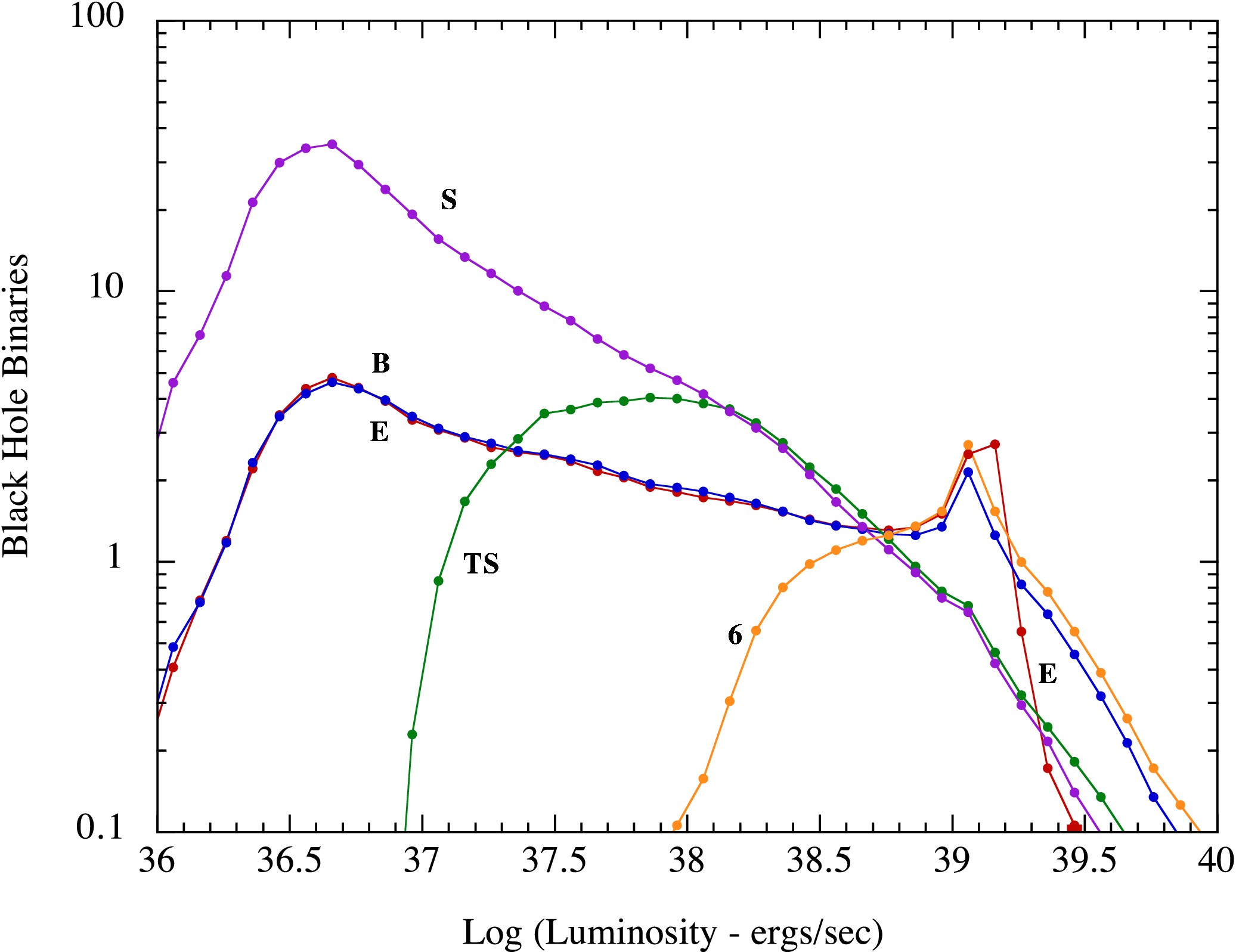}    
\caption{Luminosity functions for a galaxy undergoing {\em continuous}
star formation at a fixed age $\gg 10^8$ yr.  The core-collapse SN
rate is taken to be 0.01 yr$^{-1}$.  The results are derived from
Fig.\ \ref{fig:4panel} by weighting each column of pixels by the time
interval represented by that pixel, and then summing each row of
pixels over all columns.  Models B, E, S, TS, and 6 are defined in the
text.  \label{fig:lumfce}}
\end{center}
\end{figure}

The type of information contained in the `data' files used to produce
Fig.\ \ref{fig:4panel} can also be used to construct the black-hole
binary X-ray luminosity function for any history of star formation.
To illustrate this, we compute the luminosity function for a galaxy
undergoing continuous, uniform star formation for times exceeding
billions of years.  To do this we merely sum over the columns of the
$L_x-t_{\rm ev}$ matrix, multiplying each column by its width in time
(noting that the widths of the columns increase logarithmically with
$t_{\rm ev}$).  The results are shown in Fig. \ref{fig:lumfce} for an
assumed formation rate of black-hole binaries of $\mathcal R_{\rm BH} =
10^{-6}$ yr$^{-1}$ (PRH), which in turn utilizes a SN rate of 0.01 yr$^{-1}$ 
(e.g., Cappellaro, Evans, \& Turatto 1999).  The results in Fig.\ \ref{fig:lumfce}, 
for three of our models (B,E, and 6), indicate that a typical normal
spiral galaxy such as our own would have about a half dozen ULXs at
any given time.  Models S and TS, with a smaller population of
initially massive stars, produce of order one ULX, in closer agreement
with the observed number in our Galaxy.  

\begin{figure}
\begin{center}
\includegraphics[width=0.45\textwidth]{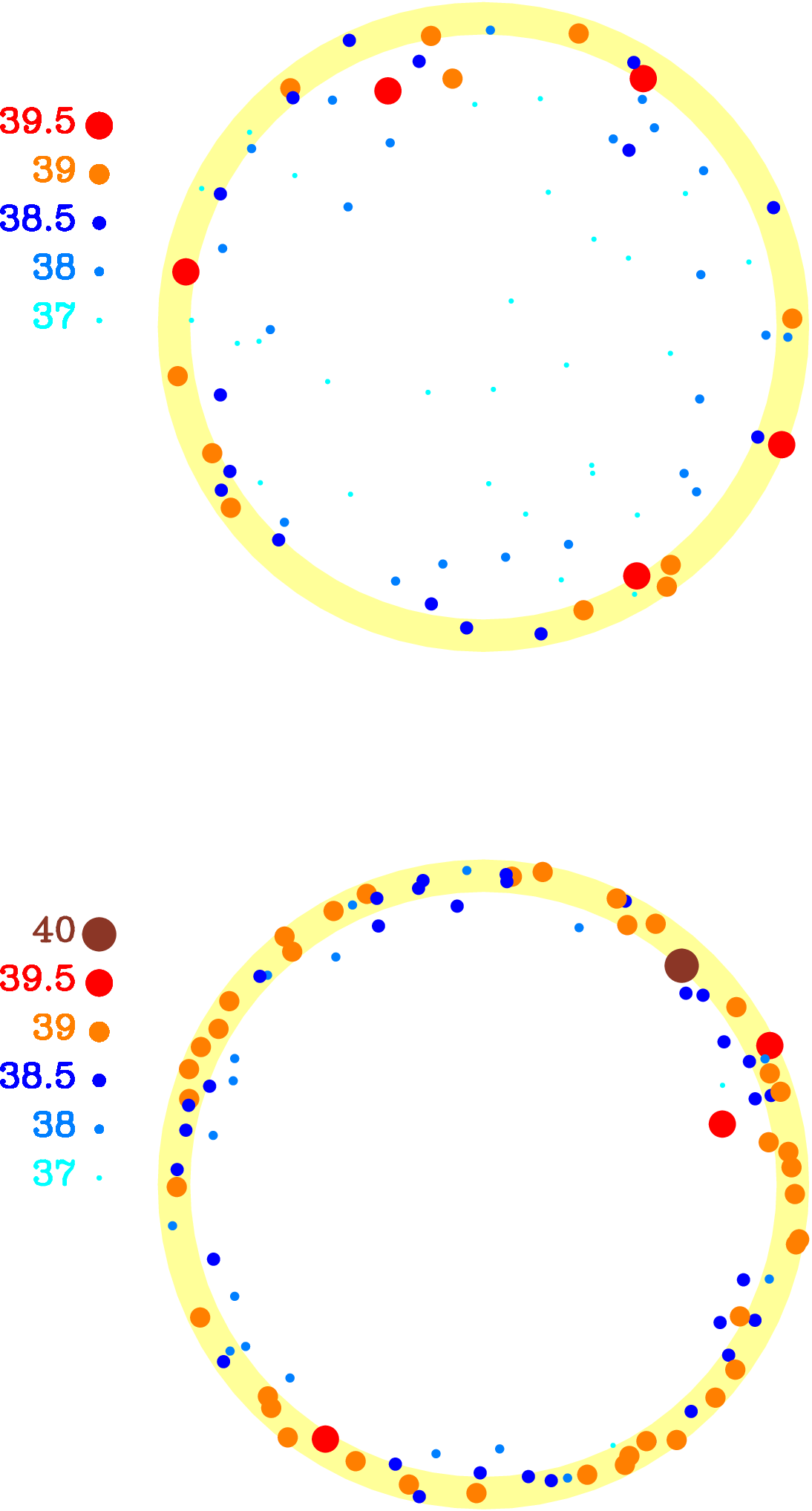}         
\caption{Simulated representation of the luminous black-hole X-ray
binaries in a galaxy where a star formation wave was triggered by a
catastrophic event at the center some $5 \times 10^8$ years ago.  The
top and bottom panels were produced for Models B and 6, respectively.
The wave of star formation is assumed to propagate outward at a
constant speed.  The surface density of interstellar gas used to form
the stars is taken to be a constant per unit surface area.  The X-ray
luminosities were chosen via a Monte Carlo technique from the
distribution for Model B (Fig.\ \ref{fig:4panel}--panel b) and Model
6.  This spatial distribution of sources is reminiscent of Chandra
images of the Cartwheel galaxy.  \label{fig:cart}}
\end{center}
\end{figure}

We can also use the results of Fig.\ \ref{fig:lumfce} to compute the 
integrated X-ray luminosity in the continuous star formation scenario
for each of our models. 
We list the results in Table 1---scaled up by a factor of 10 for better 
comparison to typical active star forming galaxies with an assumed 
core-collapse SN rate of 0.1 yr$^{-1}$.  It is instructive
to compare our results with the empirical findings of Gilfanov, Grimm, 
\& Sunyaev (2004) and Ranalli, Comastri, \& Seti (2003) who cite their 
findings in terms of the relation between the massive star formation rate 
({\em SFR}), expressed in units of $M_\odot$ yr$^{-1}$ for stars with 
$M > 5~M_\odot$, and the total X-ray luminosity $L_{\rm x,tot}$.  
To intercompare our theoretically derived results with their empirical findings, 
it is helpful to have a conversion factor between the rate of core-collapse 
SNe ($\mathcal{R}_{\rm SN}$) and the {\em SFR}, which we estimate
to be $\mathcal{R}_{\rm SN} \simeq 0.025~SFR$ for a Salpeter-type IMF.
With this conversion we find that our adopted value of $\mathcal{R}_{\rm SN}
= 0.1$ yr$^{-1}$ corresponds to an {\em SFR} of $\sim 4~M_\odot$ yr$^{-1}$.
For this particular value of {\em SFR} Gilfanov et al. (2004) and Ranalli
et al. (2003) find an average value of $L_{\rm 
x,tot} \simeq 2 \times 10^{40}$ ergs s$^{-1}$.  Our tabulated values of
$L_{\rm x,tot}$ (Table 1) for continuous star formation are typically an 
order of magnitude larger than than the empirical value.   The values
for $L_{\rm x,tot}$ in Table 1 for impulsive star formation are in better 
agreement with those of Gilfanov et al. (2004) and Ranalli et al. (2003),
but there our normalization of $10^6$ core-collapse SNe, which corresponds
to, e.g., $\mathcal{R}_{\rm SN} = 0.1$ yr$^{-1}$ for a finite time of $10^7$ years, 
probably {\em underestimates} the actual number of SNe contributing to the 
observed population of luminous X-ray sources.  On the other hand, the 
assumption of a completely continuous star formation scenario likely 
{\em overestimates} the number of luminous systems.  On balance, we 
conclude that the comparison of our results (Table 1) to those of Gilfanov 
et al. (2004) and Ranalli et al. (2003) indicates that our adopted formation 
rate of black-hole binaries of $10^{-6}$ yr$^{-1}$ for $\mathcal{R}_{\rm SN} 
= 0.01$ yr$^{-1}$ is likely too high by a factor of $\sim 3$.  This probably also
indicates that our normalization to an effective value of $\lambda = 0.1$ 
(the binding-energy parameter of the BH progenitor) is likely a bit too
large; perhaps a better selection would be $\lambda = 0.06-0.07$.

Finally, we illustrate how, for external galaxies, our results can be
used to relate the spatial offsets between ULXs and current star
forming regions.  For this we use the evolving luminosity functions
after an impulsive star formation event (e.g., Fig.\
\ref{fig:4panel}).  In this exercise we make a crude model for the
annular concentration of luminous X-ray sources in the Cartwheel
galaxy.  We envision an unperturbed spiral galaxy with a uniform
surface column density of interstellar gas as the initial condition.
We assume that some $5 \times 10^8$ years ago a smaller galaxy passed
through the center of the spiral and triggered a wave of star
formation (Hernquist \& Weil 1993) which has just now reached a radius
marked by numerous HII regions and luminous X-ray sources.  We choose
random locations over the surface of the disk (per unit area); the
radial distance of each location then specifies the time since the
star formation wave has passed.  The results shown in Fig.\
\ref{fig:4panel}, panel (b) are then used as a probability
distribution to choose an X-ray luminosity.  This is repeated until
100 X-ray sources have been chosen for the realization.  The results
are shown in the top panel of Fig.\ \ref{fig:cart}.  The yellow ring
marks the outer 10\% in radius.  Note that while the luminous X-ray
sources are indeed concentrated in the outer ring, there are still a
number of luminous sources ($\ga 10^{38}$ ergs s$^{-1}$) well inside the 
ring.  We then repeated this exercise for Model 6 where initially more 
massive donor stars are emphasized; for this model the sources are even 
more concentrated in the outer 10\% of the simulated galaxy (lower panel in 
Fig.\,\ref{fig:cart}).
    
\section{Discussion and Conclusions}

\begin{figure}
\begin{center}
\includegraphics[angle=-90,width=0.47\textwidth]{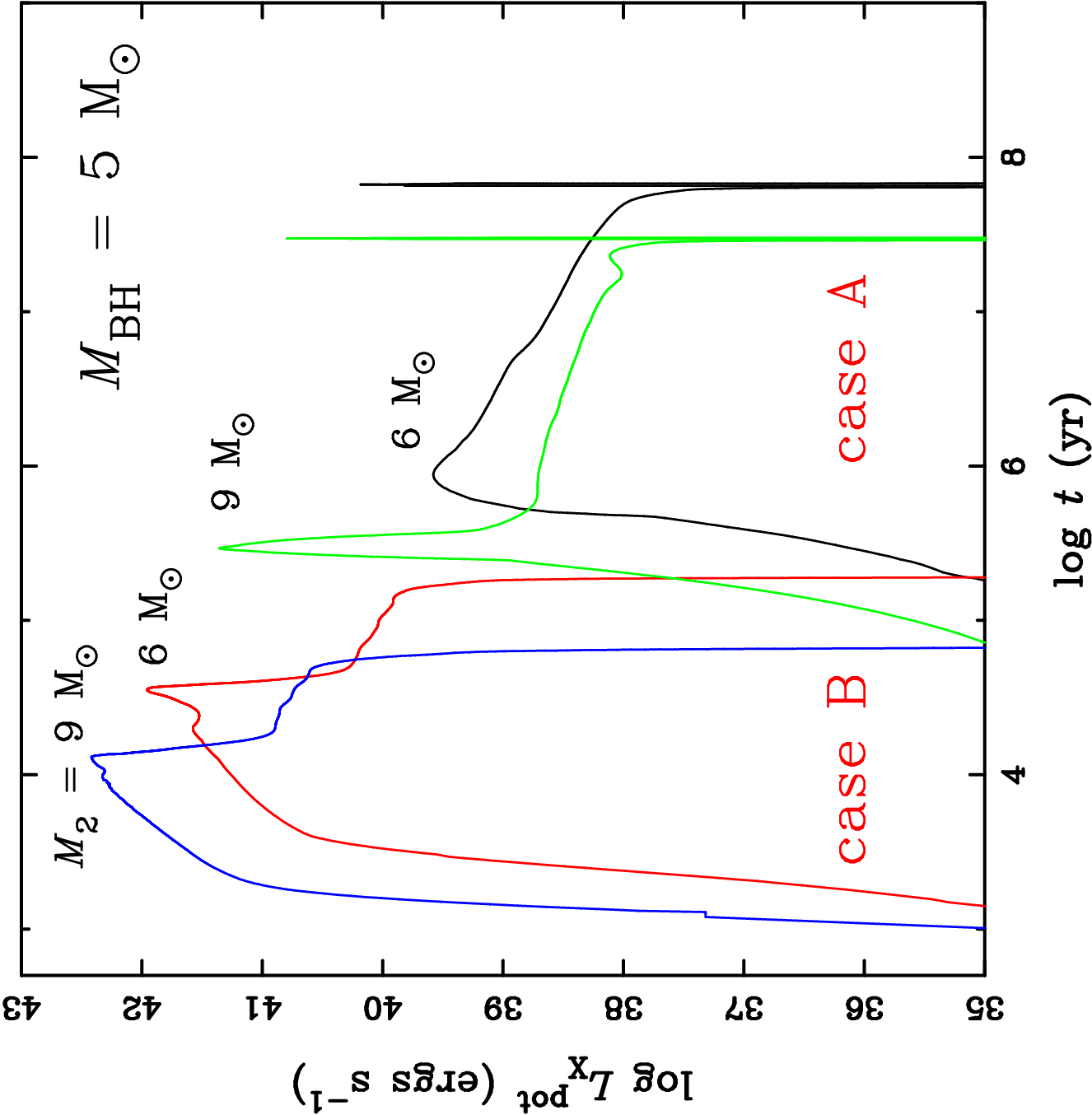}
\caption{Potential X-ray luminosities as a function of time since the 
beginning of mass transfer (offset by $10^3$ yr for clarity) for 4
black-hole X-ray binary evolution sequences. Each color corresponds to
a different mass or evolutionary state of the donor star when mass
transfer commences (as indicated).  In all cases the initial mass of
the black hole is $5~M_\odot$.  The black and red curves are for donor
stars with initial masses of $6~M_\odot$, while the green and blue
curves are for donor stars with initial masses of $9~M_\odot$.  For
the black and green curves the donor star commences mass transfer in
the middle of the main sequence (case A mass transfer).  The red and
blue curves correspond to the case where the donor star has evolved
off the main sequence and traverses the Hertzsprung gap at the start
of mass transfer (early case B mass transfer).  In all four
calculations, thermal timescale phases last of order $10^5$ yr.  Since
these involve somewhat lower-mass donor stars than the main
contributors to our ULX candidates, they may be more common, which can
compensate somewhat for the shorter duration of the high mass transfer
rate episodes.  
\label{fig:thermal}}
\end{center}
\end{figure}

We have computed a unique grid of 52 black-hole binary evolution
models covering donors in the mass range of $2-17~M_\odot$ and 4
different evolutionary phases at the onset of mass transfer from the
ZAMS to the TAMS.  We chose the incipient black-hole binaries from
regions of the $P_{\rm orb}-M_2$ plane that were suggestive of the
results of the binary population synthesis study by PRH.  Several
different assumptions were made in choosing the donor masses in the
incipient population of black-hole binaries (i.e., Model sets \{E,B,TB\}, 
\{6\}, and \{S,TS\}).  For each incipient binary chosen, we utilized an
evolution track that was interpolated from among the binary evolution
models in our grid.  In this way we computed luminosity functions for
large sets of simulated black-hole binaries.  The masses of all 
the black holes in our evolutionary calculations were taken to be
initially $10~M_\odot$.  In future studies the black-hole mass should
be taken directly from the population synthesis calculations, and
would then range from $\sim 5-15~M_\odot$ (based on observational
constraints; e.g., McClintock \& Remillard 2004).  The lower
black-hole masses, when paired with the higher-mass donors (e.g.,
$\ga 12~M_\odot$), would lead to dynamically unstable mass transfer
and would then not enter the population.  The inclusion of lower-mass
black holes will have some modest effect, but certainly not a dominant one, 
on the luminosity functions that we calculate (see also Fig.\ \ref{fig:thermal} 
and the associated discussion).

Our luminosity functions are sensitive to two principal input
parameters.  First, the value of $\lambda$ (related to the binding
energy of the progenitor of the black hole) directly affects the mass
distribution of donor stars in the black-hole binaries.  As discussed
in the text, there are significant uncertainties in the appropriate
values of $\lambda$ to use.  The smaller (larger) the value of
$\lambda$ the more the incipient donor stars are weighted toward
higher (lower) mass.  The contributions to the ULX population come
from, in decreasing order of importance, black-hole binaries with: (i)
high-mass companions, (ii) mid- to high-mass donors on 
the giant branch, and (iii) lower-mass companions that lead to transient
behavior.  Second, our results are sensitive to the factor by which we
allow for super-Eddington luminosities.  In most of our calculations
we have used a factor of 10 times the usual values of $L_{\rm Edd}$
(as estimated by Begelman 2002).  This substantially helps in the
production of ULXs.  However, we find that any further increase in
this enhancement factor does not significantly increase the number of
ULXs since the systems then become limited by the available
mass-transfer rates.

We find generally encouraging agreement between the ULX
populations that we are able to generate and the observations---at
least for the ULX luminosity function below $\sim 2 \times 10^{40}$ ergs 
s$^{-1}$.  Some of our models, in
particular Model B, yield substantial numbers of ULXs for times up to
30 Myr after a star formation event, and a few such sources for times
up to 100 Myr.  We have found, however, by comparing our production
rate of ULXs and the concomitant values of $L_{\rm x,tot}$ with the
empirical relation between the star formation rates and $L_{\rm x,tot}$
derived by Gilfanov et al. (2004) and Ranalli et al. (2003) that our adopted
black-hole binary formation rate of $10^{-6}$ yr$^{-1}$ for a core-collapse
SNe rate of 0.01 yr$^{-1}$ is likely too high by a factor of $\sim 3$.
Our calculated evolution of the luminosity function
after a star formation event can explain spatial offsets between the
location of ULXs and current-epoch star forming regions (e.g., Zezas
\& Fabbiano 2002; Fabbiano \& White 2004, and references therein),
provided that the wave of star formation passed from the former to the
latter.  This is a possible alternative explanation to the one which
requires that the ULXs are given natal kicks and ejected from the star
clusters in which they are born (e.g., Zezas \& Fabbiano 2002). 

	Throughout this work all of the cited luminosities in the figures and table have been bolometric.  On the other hand, most of the literature on ULXs cites $L_x$ in the $2-10$ keV band due to the sensitivity range of X-ray telescopes.   It is difficult to estimate what fraction, $f$, of the bolometric X-ray luminosity emerges in the $2-10$ keV band precisely because very few measurements extend to substantially lower or higher energies.  Nothwithstanding this limitation, we have used various spectral shapes for ULXs reported in the literature to estimate $f$.  We find values of $f$ between 0.1 and 0.5 with a most probable value of $\sim$0.3.  This range for $f$ should hold unless a large fraction of the luminosity is emitted in a low-temperature multi-component disk spectrum that falls largely outside the $2-10$ keV range.  (For a related discussion see Appendix A2 of Portegies Zwart, Dewi, \& Maccarone 2004.)  We can foresee two distinct possibilities for the factor $f$: (1) it falls within a fairly narrow range (e.g., near $\sim$0.3) for most ULXs, or (2) it varies from source to source or temporally for individual sources.  In the first instance,  the entire luminosity functions that we generate could be shifted downward in luminosity such that, e.g., the peak of the luminosity function for model B drops from $\sim 2 \times 10^{39}$ ergs s$^{-1}$ to $\sim 6 \times 10^{38}$ ergs s$^{-1}$, and the highest (and rarest) luminosity sources of $\sim 2 \times 10^{40}$ ergs s$^{-1}$ fall to $\sim 6 \times 10^{39}$ ergs s$^{-1}$.  By contrast, in the second case, if $f$ varies from low to high values (among sources or temporally) then our luminosity functions might be not be so seriously shifted to lower luminosities, but would yield fewer luminous sources overall.  This, in turn, would actually help the ``overproduction'' problem discussed in \S3.  If the net result of smaller values of $f$ is to shift our simulated luminosity functions to substantially lower luminosities, then this suggests two possibilities:  (1) the basic results we have generated are correct, but then the ``Begelman factor'' would have to be as high as $\sim$30 instead of 10 for black-hole accretors with masses of $\sim 10~M_\odot$, or (2) the most luminous ULXs with $L_x {\rm(2-10~keV)} \ga 10^{40}$ ergs s$^{-1}$ indeed represent a different class of objects, e.g., IMBHs.  More work is needed both observationally, to determine $f$ empirically with greater certainty, and theoretically, especially regarding the photon bubble instability model to ascertain if super-Eddington factors of $\sim$ 30 are possible.
  
In this work we have not included the contributions from very luminous
neutron star binaries (e.g., HMXBs resembling SMC X-1 and LMC X-4).  
Such systems are likely to be very short lived (e.g., $\sim 10^5$ yr; see
e.g., Levine et al.\ 1991; 1993; 2000) due to the tidal interactions
that drive the Roche lobe rapidly through the atmosphere of the
primary.  However, such neutron star binaries are much easier to form
than black-hole binaries, and therefore their contributions to the
luminosity function between $\sim 10^{37}$ and $5 \times 10^{38}$ ergs
s$^{-1}$ should be substantial.  In any study attempting to compute
the entire luminosity function, the luminous neutron star systems
clearly need to be included (e.g., Belczynski et al.\ 2004).  The
intention of the present study, however, was limited to evaluating the
contribution of stellar-mass black-hole binaries to the ULX
population.

One further caveat is in order with regard to the success of our
model for the ULXs.  Most of the ULXs that we generate in our
population synthesis have high-mass donors and are not transients.  By
contrast, many of the black-hole binaries discovered and studied in
our Galaxy are transients whose time-averaged mass-transfer rate is
well below Eddington and whose companions are rather low in mass 
(i.e., $\la 1~M_\odot$) (see, e.g., McClintock \& Remillard 2004).  
This apparent discrepancy between
the two populations was discussed in PRH, and hinges on the fact that
it is difficult, in the context of the present model, to produce black
hole binaries with low initial donor masses.  This problem arises due 
to the fact that a secondary of $\la 2~M_\odot$ tends to merge with the 
core of the massive black-hole progenitor instead of ejecting the common 
envelope.  Also, as demonstrated in PRH, black-hole binaries with initial 
donor masses $\ga 2~M_\odot$ {\em can} evolve into the type of compact 
systems often seen in the Galaxy---but only if there is a significant source 
of systemic orbital angular momentum loss, e.g., magnetic braking, which 
has been assumed not to operate in stars with radiative outer envelopes
(Parker 1955; Pylyser \& Savonije 1988).  We note, however, that there is a 
subset of stars with radiative envelopes, the Ap/Bp stars, with masses up to 
$\sim 3\,M_{\odot}$ that have strong magnetic fields ($\sim 10\,$kG) and 
long rotation periods (e.g., Hubrig et al.\ 2000). While the origin of the large 
magnetic fields in these stars and their relation to the rotation periods are 
presently not understood, their very existence may suggest that there could 
be a subset of stars with radiative envelopes where magnetic braking is
operative.  This is clearly a problem that warrants further consideration and 
modeling.

King et al.\ (2001) discuss the possible importance of thermal
timescale mass transfer (see, e.g., PRH) onto black holes as the
driver of high rates of mass transfer in ULXs.  We note, however, that
most ULXs in our simulations are binaries where mass transfer is
driven by the nuclear evolution of the massive donor, most commonly
hydrogen burning in the core, but in a fraction of systems ($\la
5\,$\%) by hydrogen shell burning as the donor star ascends the giant
branch. Even higher mass-transfer rates can indeed be attained during
phases where the secondary is out of thermal equilibrium, either
during thermal timescale mass transfer or when the secondary moves
across the Hertzsprung gap (so-called early case B mass
transfer). However, since the lifetimes of these phases are typically
several orders of magnitude shorter than the phases where mass
transfer is driven by the nuclear evolution, few ULXs are
expected to be found in this phase (see the dashed curve in Fig.\
\ref{fig:4bhev})\footnote{This is generally true for mass transfer
from a more massive star to a less massive star and, as is well known,
explains, e.g., the paucity of $\beta$ Lyrae systems compared to
Algol-type systems, believed to be their descendants (for a general
discussion, see Paczy\'nksi 1971).}. To illustrate this further we
have done some additional binary calculations, shown in Fig.\
\ref{fig:thermal}, for secondaries with initial masses of 6 and
9~M$_{\sun}$, respectively, in two evolutionary phases: in the middle of
the main sequence (case A) and just after the secondary has left the
main sequence and started to move across the Hertzsprung gap (early
case B). In all calculations, the initial mass of the black hole was taken
to be 5~M$_{\sun}$. In the 6~M$_{\sun}$ case A sequence, the secondary
always remains in thermal equilibrium and mass-transfer is entirely
driven by the nuclear evolution of the core. For about $\sim 10^7$~yr
its potential X-ray luminosity (just) exceeds $10^{39}$ ergs~s$^{-1}$
and may be classified as a ULX during this phase. In contrast, the
9~M$_{\sun}$ case A sequence experiences a phase of rapid, thermal
timescale transfer, lasting $\sim 2 \times 10^5$~yr. During this phase
the potential X-ray luminosity reaches a peak exceeding $10^{41}$
ergs~s$^{-1}$. After the thermal timescale phase, mass transfer
continues but now on a nuclear timescale for the next $3\times
10^7$~yr at a rate that is not sufficient to power a ULX. Both case A
systems appear as ULXs again as they ascend the giant branch where
their evolution is driven by hydrogen shell burning. In the case B
sequences, where mass transfer is driven by the thermal timescale
evolution of the secondary across the Hertzsprung gap, the
mass-transfer phases last only $\sim 10^5$~yr, but reach peak
mass-transfer rates of $10^{-4}$ and $8\times
10^{-4}$~M$_{\sun}$~yr$^{-1}$ for the 6 and 9~M$_{\sun}$ sequences,
respectively, and corresponding peak potential X-ray luminosities of
order and above $10^{42}$ ergs~s$^{-1}$. Note that in these latter
sequences, the mass-transfer rates, and hence potential X-ray
luminosities, drop by 1.5 orders of magnitude after the mass-ratio
has been reversed and the orbit starts to expand rapidly. During this
latter phase, the potential X-ray luminosities remain in the ULX
regime, but decrease gradually (the moderately sloping portions of the
curves in Fig~\ref{fig:thermal}).

These sequences (Fig.\ \ref{fig:thermal}) demonstrate that during 
thermal timescale mass-transfer
phases, very high mass-transfer rates are attained, providing
potentially enough fuel to power even the most luminous ULXs, but that
the duration of these phases, which obviously scales as the inverse
of the characteristic mass transfer rate, is short compared to the
duration of phases where mass transfer is driven by the nuclear
evolution of the secondary.  
A compensating factor is that lower-mass secondaries can in principle produce
mass-transfer rates in the ULX regime during thermal timescale phases
(onto low-mass black holes) that they could not otherwise do during the 
nuclear-driven phases.  This effect may somewhat, but not drastically increase 
the relative number of thermal timescale systems compared to nuclear timescale 
systems.

Transient source behavior, due to the thermal-ionization disk
instability (e.g., van Paradijs 1996; King, Kolb, \& Burderi 1996;
Dubus, Hameury, \& Lasota 2001) has been invoked as a way of producing
higher mass transfer rates, albeit only for concomitantly reduced
fractional times of the total binary mass transfer phase.  We have found that
when we include transient source behavior, which occurs predominantly
in systems with lower average luminosity (e.g., $\la 10^{37}$ ergs s$^{-1}$),
there is essentially no change in the luminosity function for times
$\la 100$ Myr; thereafter, there is a modest enhancement of
high-$L_x$ sources due to transient source behavior, but the bulk of
these objects have luminosities $\la 5 \times 10^{38}$ ergs s$^{-1}$
(see Fig. \ref{fig:4panel}, panel [d]). 

The giant phase of the donor star is also a natural mechanism to
consider for driving mass transfer rates that could enhance the
ULX population.  The mass transfer phase driven by the ascent of
the donor up the giant branch is dramatically evident in Fig.\
\ref{fig:14bhev} as ``spikes'' in $L_x$ near the {\em end} of the
evolution.  Indeed the mass transfer rate increases during this phase
by 1\,--\,2 orders of magnitude.  However, this phase is relatively short
lived and tends not to contribute dramatically to the population of
ULXs.  Fig.\ \ref{fig:4panel} shows the integrated contribution of the
giant branch phase of mass transfer to the ULX population---these are
seen as a red ridge lying $\sim$1.5 orders of magnitude above the peak
in the luminosity function for times $\ga 15$ Myr.  One could
argue that in galaxies where the Chandra sensitivity limit is $\sim
10^{39}$ ergs s$^{-1}$, these (giant branch) sources would be the only
ones visible at $t_{\rm ev} \ga$ 15 Myr.  This is true, but then
the issue arises as to what absolute numbers of sources are predicted.
As can be seen from Fig.\ \ref{fig:cumlumf} the contributions from
giant-branch donor sources do show up as a modestly significant
feature for $L_x \ga 10^{40}$ ergs s$^{-1}$ at $\sim$30 Myr.

\begin{figure}
\begin{center}
\includegraphics[angle=0,width=0.47\textwidth]{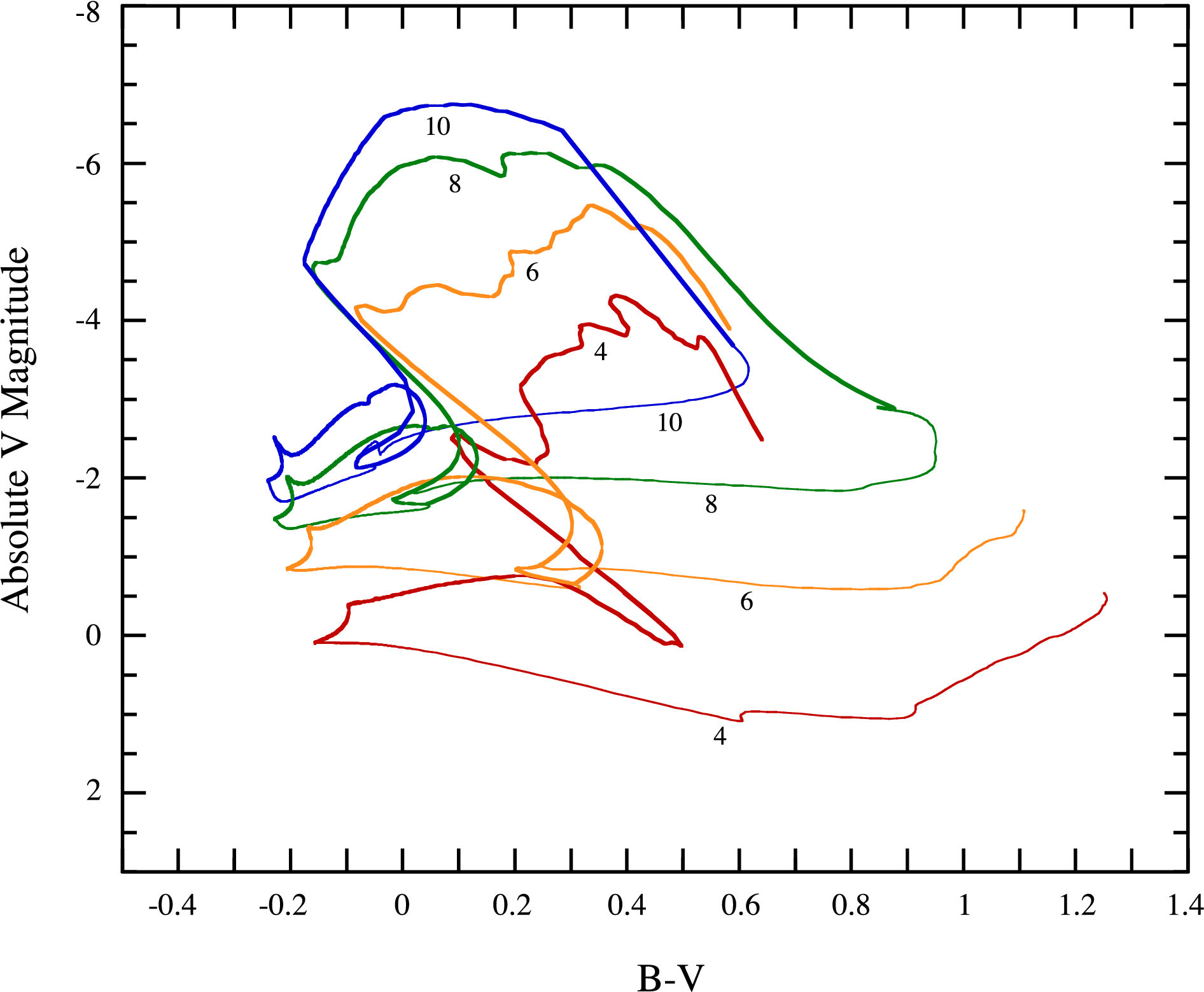}
\caption{Evolution tracks in the HR diagram are shown for 4 of our models; 
these are for donors stars that had initial masses of 4, 6, 8, and 10 $M_\odot$ 
and were on the zero-age main sequence at the start of mass transfer.  Each binary 
system is represented by two different evolution tracks; the thin curve is the 
contribution from the donor star alone, while the thicker curve (of the same color) 
describes the track for the total system light -- both disk and donor star. 
\label{fig:HRdiag}}
\end{center}
\end{figure}

Our models also have some predictive power concerning the optical 
appearance of the ULX black-hole binaries.   We have computed
tracks for our binary evolution models in the HR diagram, taking 
into account both the contribution from the donor star and the accretion
disk.  Light from the disk, in turn, results from reprocessing of X-radiation
and from viscous heating.   We computed the effective temperature of the
disk as a function of radius, $r$, from the following simple expression that
we derived:
\bea
T(r) \simeq \left(\frac{L_x}{4\pi r_{\rm min}^2 \sigma}\right)^{1/4}
\left[\frac{2}{7}\xi'x^{-12/7}(1-\alpha)+\frac{3}{2}x^{-3}\right]^{1/4} ~~,
\eea
where $x$ is the radial distance in units of $r_{\rm min}$, the inner radius of 
the disk, $\alpha$ is the X-ray albedo of the disk (which we take to be $\sim$0.7), 
and $\xi'$ is defined by $\xi' = \xi (r_{\rm min}/r_{\rm max})^{2/7}$, where, in turn, 
$\xi$ is the half thickness of the disk, $h$, at $r_{\rm max}$, in units of $r_{\rm max}$.
In this formulation $h(r)  = \xi r^{9/7} r_{\rm max}^{-2/7}$, and we take $\xi = 0.1$
as an illustrative value corresponding to a full angular thickness of the disk 
equal to $\sim 12^\circ$.  For the disk-heating problem only, we have taken 
the inner edge of the accretion disk to be at $6GM_{\rm BH}/c^2$ 
regardless of the spin of the black hole, and we have neglected the
factor $\left(1-\sqrt{r_{\rm min}/r}\right)$ found in the Shakura \& Sunyaev (1973)
solutions.  This latter approximation affects only the $x^{-3}$ term which, in any
case, contributes little to the optical flux from the disk.   The contributions to
the B and V bands from the donor stars were estimated from their bolometric
magnitude and effective temperature using conversion factors taken from Reed
(1998).  

Illustrative results for 4 evolution tracks in the HR diagram are shown in 
Fig.\ \ref{fig:HRdiag}; these are for donors stars that had initial masses of
4, 6, 8, and 10 $M_\odot$ and were on the zero-age main sequence at the start 
of mass transfer.  Each binary system is represented by two different evolution 
tracks in the HR diagram; the thin curve is the contribution from the donor star 
alone, while the thicker curve (of the same color) describes the track for the total
system light -- both disk and donor star.  In each case, it is clear that during
the early part of the binary evolution where the donor star is still on the main
sequence, the contribution from the heated accretion disk leads to a brighter
optical counterpart in the V band by about 1 magnitude.  However, during the 
short-lived giant branch phase, when the X-ray luminosity is very much larger, 
the V-band magnitudes are enhanced by $\sim$ 4 magnitudes and the system
colors are considerably bluer.  Thus, we conclude that there is a substantial 
contribution to both the color and magnitude of the optical counterparts of the 
black-hole ULXs from the accretion disk.  Moreover, we note that the contribution
of the donor star to the optical light of the system may be considerably less than
might be expected from a star of the same initial mass, especially if the mass 
transfer commences early in the evolution of the donor star.  These facts should 
be taken into account by observers trying to characterize the ``donor star'' from 
its location in an HR diagram (M. Pakull, private communication).  

Finally, we briefly discuss the issue of ULXs in elliptical galaxies
(e.g., Angelini, Loewenstein, \& Mushotzky 2001; Colbert \& Ptak 2002;
Jeltema et al.\ 2003; Ptak \& Colbert 2004).  The presence of true
ULXs with certified X-ray luminosities exceeding $10^{39}$ ergs
s$^{-1}$ would be very difficult to explain within the context our black
hole binary model unless the elliptical galaxy had a recent (e.g.,
$\la 3 \times 10^8$ yr) merger and star formation cycle.  Recently Irwin,
Bregman, \& Athey (2004) have shown, from a sample of 28 elliptical
and S0 galaxies observed with Chandra, that the number of sources with
$L_x \ga 2 \times 10^{39}$ ergs s$^{-1}$ was equal to the number
expected from background or foreground objects.  Thus, they conclude
that, with only the exception of two ULXs in globular clusters within
NGC 1399, ULXs ``are generally not found within old stellar systems".
Nonetheless, the ULXs found in early-type galaxies with $10^{39}
\la L_x \la 2 \times 10^{39}$ ergs s$^{-1}$ are not naturally explained
within the context of our model.

In summary, our calculations indicate that with a plausible set of
assumptions, a majority of the ULXs with $L_x \la 10^{40}$ ergs s$^{-1}$
in spiral galaxies can be understood
in terms of binary systems containing stellar-mass black holes.  The
systems can be evolved theoretically from the primordial binary phase
through the ULX phase---and the absolute numbers of such systems in
the populations that are computed are in agreement with the
observations to order-of-magnitude accuracy.  By contrast,
the same claims cannot yet be made for the model of ULXs that invokes
intermediate-mass black holes.  Such a model, of course, has the
advantages that (1) the Eddington limit presents no difficulties, and
(2) the spectra of some of the ULXs which exhibit low inner-disk
temperatures (see, e.g., Miller et al.\ 2003; Cropper et al. 2004) may be 
a natural consequence of more massive black holes.  However, in the 
context of the IMBH model, ideas as to how such objects are formed, 
acquire binary companions, and are ``fed'' at the requisite accretion
rates are only just beginning to be explored (see, e.g., Portegies Zwart 
et al.\ 2004; Portegies Zwart, Dewi, \& Maccarone 2004; and references
therein).

\section*{Acknowledgements}

We thank Ed Colbert, Miriam Krauss, Ron Remillard, and Tim Roberts for helpful
discussions.  One of us (SR) acknowledges support from NASA RXTE Grant
NAS5-30612.  Ph.P. thanks the {\em Chandra} subcontract to MIT for
partial support during his visit to M.I.T. EP was supported by NASA
and Chandra Postdoctoral Fellowship Program through grant number
PF2-30024.

\end{document}